\documentclass[proof]{pasj00}
\usepackage{lscape}
\usepackage{natbib}
\SetRunningHead{A.Fukui et al.}{}
\Received{}
\Accepted{}

\begin{document}

\title{Measurements of Transit Timing Variations for WASP-5b}

\author{
Akihiko~\textsc{Fukui},\altaffilmark{1,2}
Norio~\textsc{Narita},\altaffilmark{3}
Paul~J.~\textsc{Tristram},\altaffilmark{4}
Takahiro~\textsc{Sumi},\altaffilmark{1,2}
Fumio~\textsc{Abe},\altaffilmark{1,2}
Yoshitaka~\textsc{Itow},\altaffilmark{1,2}
Denis~J.~\textsc{Sullivan},\altaffilmark{10}
Ian~A.~\textsc{Bond},\altaffilmark{5}
Teruyuki~\textsc{Hirano},\altaffilmark{6,7}
Motohide~\textsc{Tamura},\altaffilmark{3}
}
\and
\author{
David~P.~\textsc{Bennett},\altaffilmark{8}
Kei~\textsc{Furusawa},\altaffilmark{2}
Fumiya~\textsc{Hayashi},\altaffilmark{2}
John~B.~\textsc{Hearnshaw},\altaffilmark{9}
Shun~\textsc{Hosaka},\altaffilmark{2}
Koki~\textsc{Kamiya},\altaffilmark{2}
Shuhei~\textsc{Kobara},\altaffilmark{2}
Aarno~\textsc{Korpela},\altaffilmark{10}
Pam~M.~\textsc{Kilmartin},\altaffilmark{4}
Wei~\textsc{Lin},\altaffilmark{5}
Cho~Hong~\textsc{Ling},\altaffilmark{5}
Shota~\textsc{Makita},\altaffilmark{2}
Kimiaki~\textsc{Masuda},\altaffilmark{2}
Yutaka~\textsc{Matsubara},\altaffilmark{2}
Noriyuki~\textsc{Miyake},\altaffilmark{2}
Yasushi~\textsc{Muraki},\altaffilmark{11}
Maiko~\textsc{Nagaya},\altaffilmark{2}
Kenta~\textsc{Nishimoto},\altaffilmark{2}
Kouji~\textsc{Ohnishi},\altaffilmark{12}
Kengo~\textsc{Omori},\altaffilmark{2}
Yvette~\textsc{Perrott},\altaffilmark{13}
Nicholas~\textsc{Rattenbury},\altaffilmark{13}
Toshiharu~\textsc{Saito},\altaffilmark{14}
Ljiljana~\textsc{Skuljan},\altaffilmark{5}
Daisuke~\textsc{Suzuki},\altaffilmark{2}
Winston~L.~\textsc{Sweatman},\altaffilmark{5}
Kohei~\textsc{Wada},\altaffilmark{11}
}
\author{(The MOA Collaboration)\altaffilmark{1}}	
\altaffiltext{1}{Microlensing Observations in Astrophysics (MOA) Collaboration, http://www.phys.canterbury.ac.nz/moa}
\altaffiltext{2}{Solar-Terrestrial Environment Laboratory, Nagoya University, Nagoya, 464-8601, Japan}

\email{afukui@stelab.nagoya-u.ac.jp}
\altaffiltext{3}{National Astronomical Observatory of Japan, 2-21-1 Osawa, Mitaka, Tokyo, 181-8588, Japan}
\altaffiltext{4}{ Mt. John Observatory, P.O. Box 56, Lake Tekapo 8770, New Zealand}
\altaffiltext{5}{Institute of Information and Mathematical Sciences, Massey University, Private Bag 102-904, North Shore Mail Centre, Auckland, New Zealand}
\altaffiltext{6}{Department of Physics, The University of Tokyo, Tokyo, 113-0033, Japan}
\altaffiltext{7}{
Department of Physics, and Kavli Institute for Astrophysics and Space Research,
Massachusetts Institute of Technology, Cambridge, MA 02139, USA}
\altaffiltext{8}{Department of Physics, University of Notre Dame, Notre Dame, IN 46556, USA}
\altaffiltext{9}{University of Canterbury, Department of Physics and Astronomy, Private Bag 4800, Christchurch 8020, New Zealand}
\altaffiltext{10}{School of Chemical and Physical Sciences, Victoria University, Wellington, New Zealand}
\altaffiltext{11}{Konan University, Kobe, Japan}
\altaffiltext{12}{Nagano National College of Technology, Nagano 381-8550, Japan}
\altaffiltext{13}{Department of Physics, University of Auckland, Auckland, New Zealand}
\altaffiltext{14}{Tokyo Metropolitan College of Industrial Technology, Tokyo 116-8523, Japan}

\maketitle

%---------------------------------------------

\begin{abstract}
We have observed 7 new transits of the `hot Jupiter' WASP-5b using a 61 cm telescope located in New Zealand, in order to search for transit timing variations (TTVs) which can be induced by additional bodies existing in the system. When combined with other available photometric and radial velocity (RV) data, we find that its transit timings do not match a linear ephemeris; the best fit $\chi^2$ values is 32.2 with 9 degrees of freedom which corresponds to a confidence level of 99.982 \% or 3.7 $\sigma$. This result indicates that excess variations of transit timings has been observed, due either to unknown systematic effects or possibly to real TTVs. 
The TTV amplitude is as large as 50 s, and if this is real, it cannot be explained by other effects than that due to an additional body or bodies. 
From the RV data, we put an upper limit on the RV amplitude caused by the possible secondary body (planet) as 21~m~s$^{-1}$, which corresponds to its mass of  22-70 M$_{\oplus}$ over the orbital period ratio of the two planets from 0.2 to 5.0.
From the TTVs data, using the numerical simulations, we place more stringent limits down to 2 M$_{\oplus}$ near 1:2 and 2:1 mean motion resonances (MMRs) with WASP-5b at the 3 $\sigma$ level, assuming that the two planets are co-planer. 
We also put an upper limit on excess of Trojan mass as 43 M$_{\oplus}$ (3 $\sigma$) using both RV and photometric data.
We also find that if the possible secondary planet has non- or a small eccentricity, its orbit would likely be near low-order MMRs.
Further follow-up photometric and spectroscopic observations will be required to confirm the reality of the TTV signal, and results such as these will provide important information for the migration mechanisms of planetary systems.
\end{abstract}


%---------------------------------------------
\section{Introduction}
\label{sec:introduction}

More than 400 extrasolar planetary systems have been found to date by several techniques, including pulsar timing, radial velocity (RV), transit, microlensing, direct imaging, and astrometry.
Among them, more than 40 systems have been revealed to have multiple planets, most of which have been detected by the RV technique \citep{2009ApJ...693.1084W}. 
Considering observational limits, multi-planetary systems are thought to be common---that is naturally expected from the standard planetary formation mechanism, called core accretion models \cite[e.g., ][]{1981PThPS..70...35H, 1996Icar..124...62P, 2002ApJ...581..666K}. Increasing the number of multi-planetary systems and studying their nature are important to improve our understandings of planetary formation mechanisms and dynamics of planetary systems.

Searching for transit timing variations (TTVs) from a constant period ephemeris is another method for identifying multi-planetary systems.
If a transiting planet is the one and only body orbiting its host star, its orbital period should be constant. On the other hand, if another perturbing body exists in the system, the orbital period will no longer be constant. 
Therefore we can find additional planets by probing TTVs, even if they do not transit their host star
 \citep{2005MNRAS.359..567A, 2005Sci...307.1288H}.
Furthermore, if they also transit, one can solve for their mass and orbital elements without RV measurements --- at least in principle.  
Not only additional planets but also exomoons \citep{2009MNRAS.392..181K} or Trojans \citep{2007ApJ...664L..51F} can be detected via the TTV method. TTVs can also be caused by other effects, such as 
orbital precession due to effects of the general relativity,  spin-induced stellar quadrupole moment, and tidal deformations of the star and the planet \citep[e.g.,][and references therein]{2010arXiv1007.2780I},
as well as variations in the quadrupole moment driven by stellar activity known as the Applegate effect \citep{1992ApJ...385..621A, 2010MNRAS.405.2037W} and orbital decay due to tidal dissipation \citep{2009Natur.460.1098H}.
However, the impact of these effects on the orbital period is relatively tiny compared to that due to additional orbiting bodies.

The TTV method is especially sensitive to planets in orbits near mean motion resonances (MMRs) with transiting planets. Even a  sub-Earth mass planet in low-order MMR is detectable using relatively small-size (0.5-1m) ground-based telescopes \citep{2007arXiv0704.0632S}. It is difficult for these telescopes to detect such low-mass planets by other methods, except via the microlensing technique \cite[e.g.,][]{2009ApJ...703.2082Y}.
Such small telescopes are relatively plentiful and widely distributed around the world and thus their availability for observing phenomena such as multiple transit timings is more flexible than that for larger aperture telescopes.

Searching for planets in MMRs with hot Jupiters is also important. According to the core accretion models, hot Jupiters are predicted to form at $\sim$ several AU where solid material is abundant enough to become a massive core and to accrete gas,  and then migrated inward to their current positions in some way; mainly either via gravitational disk-planet interaction models \citep[hereafter disk-planet interaction models, e.g.,][]{1986ApJ...309..846L}, or via planet-planet scattering and/or Kozai process followed by tidal evolution \citep[hereafter planet-planet scattering models, e.g.,][]{2008ApJ...678..498N}.
The disk-planet interaction models predict that two planets including a gas-giant planet in a system will be easily captured in low-order MMRs during their migration processes \cite[e.g.,][]{2005ApJ...626.1033T}. 
On the other hand, the planet-planet scattering models are thought to be difficult to form planets captured in MMRs.
The current distribution of the sky-projected angle between the stellar spin and planetary orbital axes measured via the Rossiter-McLaughline effect shows that a significant amount of hot Jupiters are misaligned \cite[e.g.,][hereafter T10]{2009PASJ...61L..35N, 2009ApJ...700..302W, 2010arXiv1008.2353T}. This fact implies that the planet-planet scattering models might play an important role in forming hot Jupiters.
Furthermore, \citet{2010ApJ...718L.145W} pointed out that hot Jupiters around hot stars tend to have high obliquities rather than those around cool stars, and they proposed a hypothesis that most or all hot Jupiters initially have high obliquities in some way and only cool stars have damped their obliquities. If this explanation is true, planets in MMRs with hot Jupiters would be rare.
The TTV method can thus be a powerful tool to check the existence or nonexistence of such planets around hot Jupiters and to test these planetary migration scenarios.

Although the TTV method has been employed in many searches for additional planets, to date most of them have not shown any planetary signals while they have been used to put upper limits on masses of  hypothetical additional planets in the systems \cite[e.g.,][and a summary therein]{2010ApJ...714...13A}.
Recently, three systems have been reported to show plausible TTV signals;  WASP-3 \citep{2010arXiv1006.1348M}, WASP-10 \citep{2010arXiv1009.4567M}, and Kepler-9 \citep{2010Sci}.

The transiting planet WASP-5b was discovered by \citet{2008MNRAS.387L...4A} (hereafter A08) as a 1.58 $M_\mathrm{jup}$ hot Jupiter orbiting a $V$ = 12.3 (G4V) star with a period of $1.63$ days. \cite{2009A&A...496..259G} (hereafter G09) have conducted high-precision photometric and spectroscopic follow-up observations for this system using the 8.1 m Very Large Telescope (VLT). Although the resulting photometric data suffered from uncorrectable systematic errors and were not used for their analysis, they reanalyzed photometric and RV data presented in A08 and found marginal non-zero eccentricity ($\sim 2\sigma$) for the planetary orbit. 
They also found that the reduced-$\chi^2$ of a linear fit for four transit timing data was 5.7, which might be a sign of additional bodies. 
 \citet{2009MNRAS.396.1023S}(hereafter S09) presented two high quality photometric transit light curves obtained with the 1.54 m Danish telescope and with residual standard deviations relative to theoretical fits of 0.50 mmag and 0.59 mmag. They argued that the inconsistency with a linear ephemeris could be associated with a relatively poor transit light curve from the Faulkes Telescope South (FTS).
T10 presented 33 new RV data obtained with the HARPS spectrograph, including data during a transit, and 5 CORALIE data in addition to the 11 given by A08. They derived the sky-projected spin-orbit angle of $\lambda={12.1^{\circ}} ^{+8.0^{\circ}}_{-10.0^{\circ}}$, which is consistent with a spin-orbit alignment that is naturally expected from the disk-planet interaction models for the migration mechanism of WASP-5b.
However, the planet-planet scattering models followed by the stellar obliquity dumping could still be an alternative scenario, because the host star is a ``cool star" in the context of \citet{2010ApJ...718L.145W}, which might therefore easily dump the stellar obliquity. 
We cannot exclude this scenario at this point in time,
and therefore searching for additional plants in low-order MMRs via the TTV method is useful to test these migration scenarios.

In this paper, we present observations of 7 new transits of WASP-5b obtained with a 61 cm telescope. 
We also gathered available photometric and RV data and performed joint-fit to refine transit parameters and to search for TTV signals.
In section~\ref{sec:obsandreduction} we describe our observations and data reduction. We show our light curve modeling in section~\ref{sec:modeling} and discuss results in section~\ref{sec:TTV}. We summarize our findings in section~\ref{sec:conclusion}.

%---------------------------------------------
\section{Observations and Data Reduction}
\label{sec:obsandreduction}

\subsection{Transit Observations}
\label{sec:obs}

  We observed 7 transits of WASP-5b (corresponding transit epochs, $E$, are: 160, 244, 432, 451, 459, 607, and 615 based on the ephemeris given by A08)  using the 61 cm Boller and Chivens  (B\&C) telescope located at Mt.\ John University Observatory, operated by the University of Canterbury at Lake Tekapo in New Zealand.
 The B\&C telescope is normally used for the microlensing followup program of Microlensing Observations in Astrophysics collaboration \cite[MOA;][] {2001MNRAS.327..868B, 2003ApJ...591..204S, 2010ApJ...710.1641S}, and the transit observations were obtained during low-priority observing time.  
   A 1k $\times$ 1k CCD camera of Apogee ALTA U47 is mounted on the telescope which has a focal length of 8.3 m. The CCD pixel size of  13 $\mu$m $\times$ 13 $\mu$m therefor leads to a pixel scale of \timeform{0".33} per pixel and a field of view of \timeform{5'.5} $\times$ \timeform{5'.5}. The CCD frame readout time is  5 s and has a readout noise, $n_\mathrm{read}$, of 11.5 electrons per pixel. The dark current noise, $n_\mathrm{dark}$, is 0.03 ADU per pixel per second under typical conditions. A Bessell $I$ filter was used for all observations and the exposure times were 30 s for $E$=245 and $E$=607, and 60 s for the others. 
  All images were taken with the telescope properly focused, except for the transit of $E$=459 when the telescope was slightly defocused so that the FWHM of the PSF was 9-12 pixels (\timeform{3".0}-\timeform{4".0}). The typical seeing at the observatory was $\sim\timeform{2".0}$.
An observing log is shown in table~\ref{tbl:obslog}.

\subsection{Data Reduction}
\label{sec:datareduction}

All images were initially bias subtracted and flat-field corrected in the standard manner.
We then performed synthetic aperture photometry for the target star WASP-5 and some (2 -- 4) comparison stars in the same field of view, using the following procedure.
Note that each star to be measured is enough isolated from the closest star not to be contaminated during the procedure.

  First, for each image $j$ and using a step size of 0.1 pixels, we searched for an initial photometric aperture radius $r_{\mathrm{opt,} j}$ that maximized the signal to noise ratio of the target flux. 
The signal, $F_\mathrm{star}$, is flux from the target which is equal to the total flux in the aperture minus the corresponding sky flux, $F_\mathrm{sky} = m_\mathrm{pix} f_\mathrm{sky}$. Here, $m_\mathrm{pix}$ is the total number of pixels in the aperture and $f_\mathrm{sky}$ is the median flux per pixel in an annulus between radii $r_{\mathrm{opt,} j} + 50$ and $r_{\mathrm{opt,} j} + 60$ where the flux contribution from the target star is negligible.
%The signal, $S$, is equal to the total target flux in the aperture minus the corresponding sky flux, which was calculated from the median flux in an annulus between radii $r_{\mathrm{opt,} j} + 50$ and $r_{\mathrm{opt,} j} + 60$ where the contribution of target flux is negligible. 
The total noise contributing to the signal is modeled as:
\begin{equation}
\label{eq:noise}
N_\mathrm{total} = \sqrt{ N_\mathrm{star}^{2} + N_\mathrm{sky}^{2} + N_\mathrm{read}^{2} + N_\mathrm{dark}^2 + N_\mathrm{scin}^2 },
\end{equation}
where $N_\mathrm{star} = \sqrt{F_\mathrm{star}/g}$ is the photon noise of the target star, $N_\mathrm{sky} = \sqrt{F_\mathrm{sky}/g}$ is the sky background noise,  $N_\mathrm{read} = \sqrt{m_\mathrm{pix}} n_\mathrm{read}/g$ is the read out noise, $N_\mathrm{dark} =  \sqrt{m_\mathrm{pix}} n_\mathrm{dark} \Delta t$ is the dark current noise, and
%where $g$ is the CCD gain [electron ADU$^{-1}$], 
%$N_\mathrm{sky}$, $N_\mathrm{read}$, $N_\mathrm{dark}$, and
 $N_\mathrm{scin}$ is the atmospheric scintillation, respectively. Here, $g$ = 1.19 electron per ADU is the CCD gain  and $\Delta t$ is the exposure time in seconds. $n_\mathrm{read}$ and $n_\mathrm{dark}$ are defined in the previous section.
% are noises from the star flux, the sky background, the readout noise, the dark current, and the atmospheric scintillation respectively. 
The scintillation noise can be expressed as 
\begin{eqnarray}
\label{eq:noise}
N_\mathrm{scin} = \sigma_\mathrm{0} \frac{z^{7/4}} {D^{2/3}(\Delta t)^{1/2}} \exp \left( -\frac{h}{8000 \mathrm{m}} \right) F_\mathrm{star},
\end{eqnarray}
where $z$ is the airmass, $D$ is the telescope diameter in cm, and $h = 1029$ m is the observatory altitude \citep{1998PASP..110..610D,1967AJ.....72..747Y}. $\sigma_\mathrm0$ is a coefficient which is often taken to be 0.064 and we also adopted this value.
At typical conditions with exposure time of 60 s, the main contributor to the total noise was the photon noise ($\sim$0.0012\% of the target flux) and the contributions from the other noises were one or two orders of magnitudes less than that.

  Next, after determination of the $r_{\mathrm{opt,} j}$ value for the target star, this radius was applied to all other comparison stars on the same image to measure their fluxes.
  The target flux was then normalized to a reference flux that was created as the weighted average of the fluxes of the comparison stars.
%Initial light curves for WASP-5 were created from the target fluxes normalized to a reference flux that was calculated from the weighted average of the fluxes of the comparison stars. 
The photometric error of the normalized flux was initially estimated using the equation (\ref{eq:noise}) and the error propagation equation. 

Finally,  for each transit $E$, we searched for a scaling factor $\gamma_E$ (in the range 0.7 to 1.50) to the initial radius $r_{\mathrm{opt,} j}$  in order to  minimize the rms scatter of the derived out of transit (OOT) light curve (either before or after transit).

  In order to assess  the validity of our method, we also performed aperture photometry with the fixed radius $r_{\mathrm{fix,} E}$ for each transit which was selected so as to produce the minimum rms scatter for the OOT light curve. The result of this exercise was that the OOT rms values derived from the $\gamma_E r_{\mathrm{opt,} j}$ apertures were slightly better than those from the $r_{\mathrm{fix,} E}$ apertures for all transit light curves, except for the transit $E$=615 for which there are a relatively small  number of OOT data points, possibly leading to a large statistical fluctuation. Table \ref{tbl:scatters} shows the comparison of OOT rms values derived from the two methods.
 In order to maintain consistency, all light curves were generated by using the scaled apertures method and these were used for further analyses.

All time stamps of observations, which were recorded in the FITS headers  as the observation start and end points in units of  Julian Day (JD) based on  Coordinated Universal Time (UTC) and obtained from the network time protocol, were converted to the mid-observation points in units of  Barycentric Julian Day (BJD) based on Barycentric Dynamical Time (TDB)  using the code {\tt UTC2BJD}  \citep{2010arXiv1005.4415E}. The time given the network time protocol has occasionally been confirmed to coincide with the GPS time within 1 s  by observers. The time-conversion error does not exceed 20 ms.

%---------------------------------------------
\section{Light Curve Modeling}
\label{sec:modeling}

\subsection{Systematic correction}
\label{sec:syscorr}

Since apparent systematic trends were seen in the derived light curves, we corrected these systematic effects as follows. 
The systematic trends could arise from the changing airmass, variations in the atmospheric extinction coefficient, slow brightness variability of the target or comparison stars, and so on. 
Using the correction factor $\Delta m_\mathrm{corr}$ in the magnitude scale, 
the corrected flux $F_\mathrm{corr}$ can be expressed as
\begin{eqnarray}
\label{eq:Fcorr}
F_\mathrm{corr} = F_\mathrm{obs} \times 10^{-0.4 \Delta m_\mathrm{corr}}
\end{eqnarray}
where $F_\mathrm{obs}$ is the observed flux of the target normalized by the reference flux.
If we assume that amplitudes of the intrinsic stellar variability and the change in the second-order atmospheric extinction coefficient are both proportional to time,
$\Delta m_\mathrm{corr}$  can be expressed as
\begin{eqnarray}
\label{eq:m_corr}
\Delta m_\mathrm{corr} =  k_\mathrm{0} + k_\mathrm{z} z + k_\mathrm{t }t + k_\mathrm{tz} t z
\end{eqnarray}
where $z$ is the airmass, $t$ is the time, and  $(k_\mathrm{0}, k_\mathrm{z}, k_\mathrm{t}, k_\mathrm{tz})$ are the relevant coefficients. This equation is similar to equation (1) of \citet{2009AJ....137.3826W}, although here we use four coefficients instead of three (see Appendix). 
Provided there were enough data points, only the OOT light curve segments were fitted to estimate the systematic-correcting function.
However, three of the light curves ($E$=244, 432, and 607) did not have an adequate  amount of data points in their OOT sections,  and
also apparent systematics were seen in the first half part of the light curve of $E$=451
transit which could not be corrected if only the OOT data were used. 
Therefore, we decided to fit each light curve including the transit section while fixing the parameters defining the transit shape to the values  derived by S09, who had obtained two  transit light curves with higher precision than ours. 

\begin{figure}[ht]
\begin{center}
\FigureFile(160mm,100mm){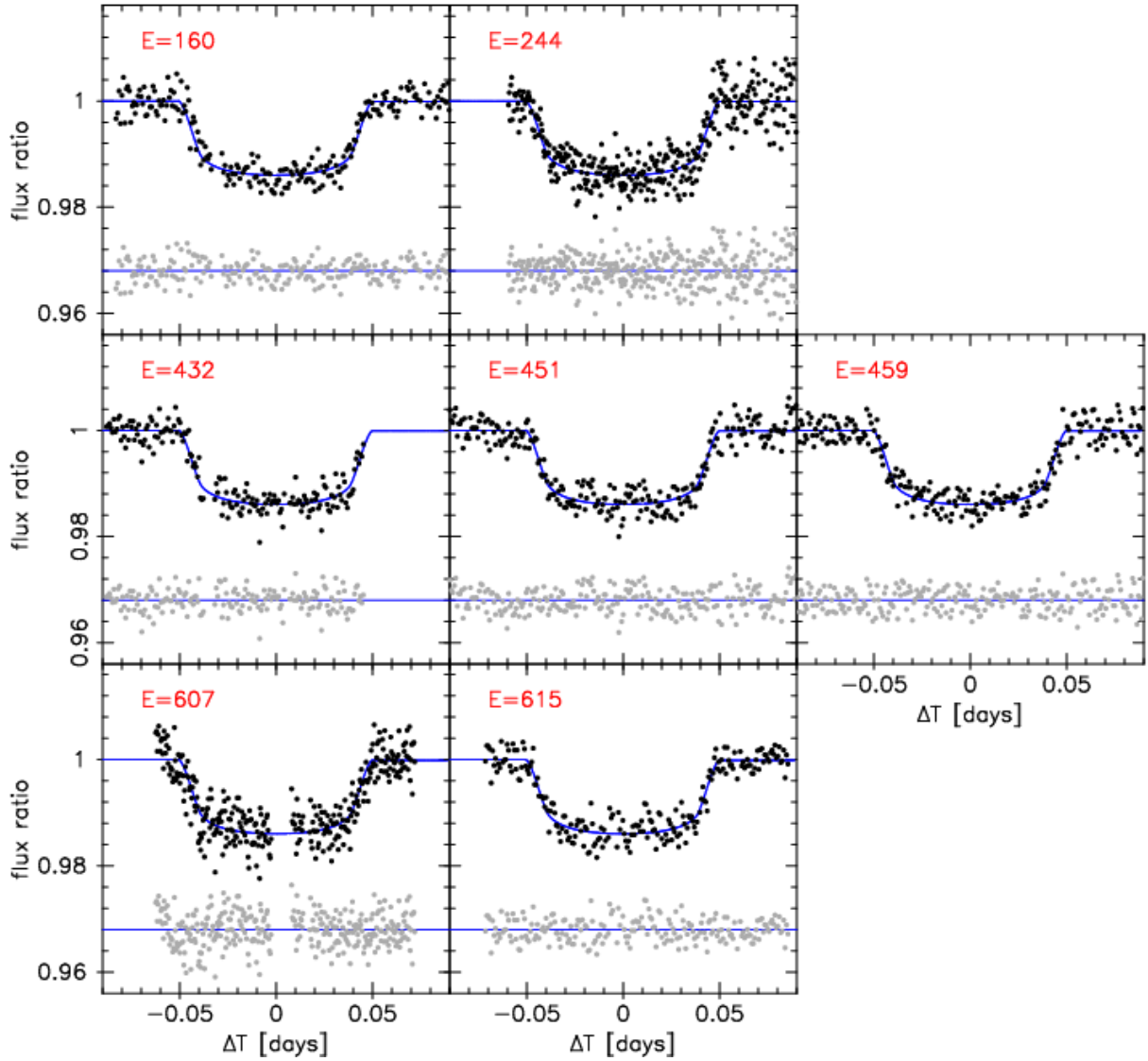}
\end{center}
\caption{
Systematic-corrected light curves of WASP-5b (black points) and their residuals (gray points) from the best fit model (solid lines) which was determined by the simultaneous fit for seven transit light curves with their normalized error bars taking red noises into account. 0.968 is added to each residual light curve for display. $E$ numbers stand for the corresponding transit epochs. 
\label{fig:multi-lcs}}
\end{figure}

To create the parameterized transit light curve, an analytic model of \citet{2002ApJ...580L.171M} was used.
This model requires 6 parameters:  the planet/star radius ratio $R_\mathrm{p} / R_\mathrm{*}$, the planetary orbit semi-major axis in units of the star radius $a/R_\mathrm{*}$, the orbital inclination to the line of sight $i$, the orbital period $P$, and two stellar limb-darkening coefficients $u_\mathrm{1}$ and $u_\mathrm{2}$. We used a quadratic limb darkening law,
\begin{eqnarray}
I(\mu) = 1 - u_\mathrm{1}(1-\mu) - u_\mathrm{2}(1-\mu)^2,
\end{eqnarray}
where $I$ is the intensity and $\mu$ is the cosine of the angle between the line of sight and the line from the center of the star to a position of the stellar surface.
For correcting systematics, these parameters were fixed to the values of 
$R_\mathrm{p} / R_\mathrm{*}$ = 0.1110, $a / R_\mathrm{*}$ = 5.4142, $i$ = 85.8$^\circ$, $P$ = $1.6284246$ d (adopted from S09), 
$u_\mathrm{1}$=0.257, and $u_2$=0.341 \citep[adopted from the tables of][]{2000A&A...363.1081C}.
Using this model and the equation (\ref{eq:Fcorr}) and (\ref{eq:m_corr}), each light curve was fitted using the $\chi^2$ minimization method with five free parameters: $k_\mathrm{0}, k_\mathrm{z}, k_\mathrm{t}, k_\mathrm{tz}$, and the mid-transit time $T_{\mathrm{c}, E}$. 
The corrected light curves are shown in figure \ref{fig:multi-lcs}.

\begin{figure}[ht]
\begin{center}
\FigureFile(120mm,100mm){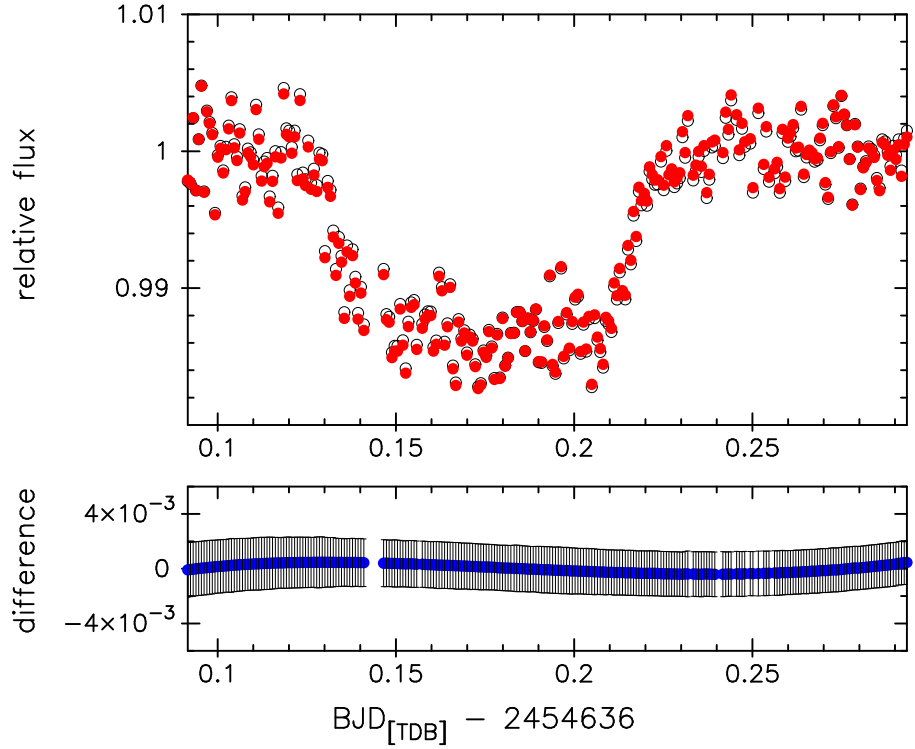}
\end{center}
\caption{
The difference (lower panel) between the light curve corrected using coefficients derived by fitting only OOT part (open circles in upper panel) and the one corrected using coefficients derived by fitting overall light curve with the transit-model parameters fixing at the values presented in S09 (filled circles in upper panel). The initial error bars estimated from the equation (1) are overplotted on the differential data.
\label{fig:oot_vs_lit}}
\end{figure}

To check for the validity of this correction, we alternatively corrected the $E$=160 complete transit light curve  by fitting only its OOT parts, then compared the two corrected light curves.
Figure \ref{fig:oot_vs_lit} shows the two light curves and their differences.
The difference was no larger than 0.00048, which is well within the typical error bar that was estimated from equation (\ref{eq:noise})  ($\sim$0.002).
Since we obtained similar results for two other complete transits ($E$=459 and 615), we decided to use the light curves corrected by the former method for further analyses.

Next, to estimate a statistical weight for each light curve, we fitted each one separately and rescaled the initial error bars so that each reduced-$\chi^2$ became unity. We also estimated time-correlated systematic noise (red noise) by using "time-averaged" method according to \citet{2008ApJ...683.1076W}.
In this method,  an error scaling factor $\beta$ ( $>1$) is calculated for each light curve by analyzing the residuals from the  best-fit model.
 If the red noise is not significant, the standard deviation after binning the residuals into $M$ bins of $N$ points, $\sigma_N$, would be expected to be
\begin{eqnarray}
\sigma_N = \frac{\sigma_1}{\sqrt{N}} \sqrt{\frac{M}{M-1}},
\end{eqnarray} 
 where $\sigma_1$ is the standard deviation of the un-binned residuals. The actual measured value of $\sigma_N$  is usually larger by  a factor $\beta$. Since the value determined for $\beta$ depends on the choice of $N$ and $M$, we adopted the maximum $\beta$ value in the range from $N=5$ to 30 and rescaled each error bar by this value. The derived values of $\beta$ and entire rms scatters of residuals are shown in table \ref{tbl:beta}.

\subsection{Joint Fit}
\label{sec:fit}

To assess the quality of our light curves compared to the previous work, 
we fitted the seven transit light curves from the B\&C telescope simultaneously  and compared the resultant parameters to the values determined by S09 and T10, in which independent data sets were used.
At this time we used the same parameterization as previous subsection and used formulae of \citet{2009ApJ...690....1O} for creating a transit light curve model.
We used common parameters of $i$, $R_\mathrm{p}/R_*$, $a/R_*$,$u_1$, and $u_2$ to all transit light curves, while used respective transit-mid times, $T_{\mathrm{c}, E}$. 
We allowed the parameters of $i$, $R_\mathrm{p}/R_*$, $a/R_*$, $u_1$, and each $T_{\mathrm{c}, E}$ to vary, while $u_2$  was fixed at the theoretical value of 0.321.
We fixed $u_\mathrm{2}$ in the fitting process since the $u_1$ and $u_2$ quantities are strongly correlated and could not both be well determined from the light curve fitting. 
The eccentricity was also fixed to zero.

We then derived the best fit parameters by minimizing the $\chi^2$ statistics 
using the {\tt AMOEBA} algorithm \citep{1992nrca.book.....P} and estimated uncertainties using the $\Delta \chi^2=1.0$ criterion, following \citet{2007PASJ...59..763N}. The resultant parameters and their 1-$\sigma$ uncertainties are shown in table \ref{tbl:fit_params}. 
The parameters from the B\&C are consistent with the ones derived by S09 and T10  within their error bars, except for the period $P$, which may be a sign of the TTVs (discussed in section \ref{sec:TTV}).

\begin{figure}[ht]
\begin{center}
\FigureFile(160mm,100mm){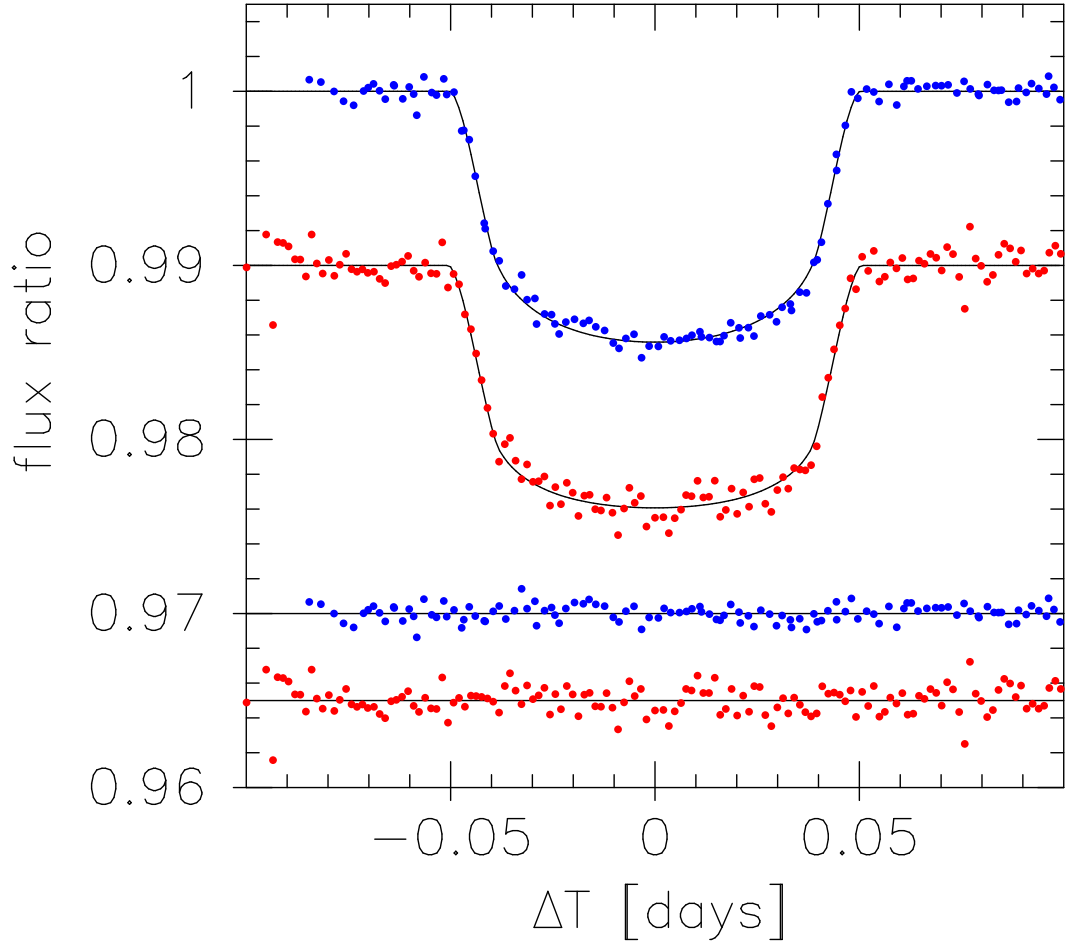}
\end{center}
\caption{Phase folded, 120-s binned light curves of WASP-5b from the Danish telescope (top) and the B\&C telescope (second) are shown. Black lines show the best-fit models for the Danish data (results for the case  using the quadratic limb-darkening law presented in S09) and for the B\&C data (results for a joint fit to the seven light curves). Third and bottom plots show their residuals (0.965 and 0.970 are added, respectively,  for display). The rms scatters are 0.00046 and 0.00081, respectively.
\label{fig:BC_vs_Sou}}
\end{figure}

 Figure \ref{fig:BC_vs_Sou} compares phase folded, 120-s binned light curves from the B\&C and the Danish (online data of S09) telescopes. Before folding the Danish data, the quoted errors were normalized and rescaled using the same method  for  the B\&C data descried in the previous section, i.e., normalized  so that the reduced-$\chi^2$ for the best-fit to each transit was unity and rescaled by a factor $\beta$ accounting for the red noise; we then took the weighted average and its error for each binned data set.
 
The rms residuals from the best-fit models for the Danish data and the B\&C data are 0.00046 and 0.00081 respectively. However, uncertainties in the derived transit-model parameters from the B\&C data are comparable to the ones from the Danish data presented in S09 (see table \ref{tbl:fit_params}).
This result initially appears surprising given the difference in the fit residuals; 
however, we found that 
 the values of the error-rescaling factor $\beta$ for the Danish data are larger than the most of that for the B\&C data (see table \ref{tbl:beta}).
 As a consequence, the average values of the binned errors of the Danish and B\&C data sets (0.00061 and 0.00084 respectively) are more comparable.
Moreover, the $R$-band filter used for the Danish observations would have resulted in
a larger limb-darkening effect than that seen using the $I$-band filter employed in the B\&C observations. This leads to a relatively poorer determination of the model parameters for the Danish data \cite[e.g.,][]{2008MNRAS.390..281P}.
In addition, S09 included limb-darkening model dependencies in their error estimations which enlarged the errors of the transit-model parameters. However, all general limb-darkening models produce a symmetrical transit shape, which has little effect on errors of transit timings. For this reason, we do not include model dependencies derived from limb-darkening in this paper.

To refine transit model parameters and to revise the timings of the previous transits, we assembled all available photometric and RV data in addition to our own data, and fitted them jointly.
The photometric dataset consists of 11 transits including one transit from the 1.2 m Euler telescope (133 data points) and one transit from the 2.0 m FTS telescope (335 data points) analyzed in A08 (private communication), two transits from the 1.54 m Danish telescope (174 data points) presented in S09, and seven transits from the B\&C telescope (1870 data points). The RV dataset consists of 16 and 33 data points from CORALIE and HARPS instruments, respectively, presented in T10.
In these available datasets, all time stamps are provided either in the form Heliocentric JD (HJD) (Euler, FTS, and Danish) or BJD (CORALIE and HARPS). However, time-standards (e.g. UTC or TDB) on which these time systems are based are unspecified in the publications.
As \citet{2010arXiv1005.4415E} alerted recently, specifying the time-standard is important in order to achieve 1 minute accuracy, and TDB-based BJD is the recommended time system.
We confirmed (in private communications) that all time systems in the available data were based on UTC , and we therefore converted all of them to the TDB-based BJD.

In order to treat the acquired photometric data equally with the B\&C data, their quoted errors were normalized and rescaled in the same manner as for the B\&C as explained in the previous subsection. The derived value of $\beta$ and the rms scatter of the residuals for each transit are shown in table \ref{tbl:beta}. 
For the RV data, systematic errors of $\sim$7~m~s$^{-1}$ may be presented (A08 and T10) in addition to the quoted internal errors, these often come from stellar activity (stellar jitter). 
To add statistical weight to the RV data,
we rescaled the quoted RV errors for both CORALIE and HARPS by adding 7~m~s$^{-1}$ quadratically. This value is consistent with the upper value from an empirical model of \citet{2005PASP..117..657W} for a G-type star.

We then fitted these data with three additional parameters: the RV semi-amplitude $K$, the sky-projected stellar rotational velocity $V \sin I_\mathrm{s}$,  and the sky-projected spin-orbit alignment angle $\lambda$.
The last two parameters were needed to model the Rossiter-McLaughline(RM) effect during transits of WASP-5b. The RM formula we used was based on \citet{2010ApJ...709..458H}, which is more appropriate than the old one used in the previous work (T10).
We allowed the parameters of $i$,  $R_\mathrm{p}/R_\mathrm{*}$, $a/R_\mathrm{*}$, $T_{\mathrm{c}, E}$, $K$, $V \sin I_\mathrm{s}$, and $\lambda$ to be free. We also fitted $u_1$ for each telescope while each $u_2$ was fixed at the theoretical value \citep{2000A&A...363.1081C, 2004A&A...428.1001C}, because, as mentioned previously, it was difficult to determine unique values for both $u_\mathrm{1}$ and $u_\mathrm{2}$ simultaneously due to their strong correlation. 
As for the eccentricity, we first allowed it to be free and obtained the value of  0.002 $^{+0.010}_{-0.002}$.
This value is consistent with zero within 1-$\sigma$ and also consistent with the result of  T10 ($e < 0.0371$, 2 $\sigma$), but marginally inconsistent with G09 ($e = 0.049 ^{+0.020}_{-0.017}$). However, the latter used a smaller RV data set in comparison with us, and hence we adopted zero for the eccentricity and re-fitted the data.
The derived value of $u_\mathrm{1}$ and the fixed value of $u_\mathrm{2}$ for each telescope are listed in table \ref{tbl:limb-darkening},
and the other best-fit parameters are shown in table \ref{tbl:fit_params}.

From the combined fit, we refined some transit model parameters, $i$, $R_\mathrm{p}/R_\mathrm{*}$, and $a/R_\mathrm{*}$ against the published values while they are consistent within their error bars. 
Uncertainties in $K$ and $V\sin I_\mathrm{s}$ became larger than the ones determined by T10 because we incorporated the stellar jitter in the RV errors.
We adopted the value of the stellar mass as the one determined spectroscopically by T10, and used this to convert fitted parameters to physical parameters. The derived values are shown in table \ref{tbl:phys_params}.  
The $\chi^2$ values for the COLALIE  and HARPS datasets are 16.0 and 32.7 respectively, which are very close to the number of data points (16 and 33 respectively).

We confirmed that the projected spin-orbital angle $\lambda$ is consistent with zero ($\lambda = 7.2^{\circ}$$^{+9.5^{\circ}}_{-9.5^{\circ}}$), which was first reported by T10, by adding independent photometric data (the B\&C and Danish data) and using an improved RM formula.

 %-------------------------------------------
 \section{Results and Discussion}
 \label{sec:TTV}

\subsection{Significance of TTV signal}
\begin{figure}[ht]
\begin{center}
\FigureFile(160mm,0mm){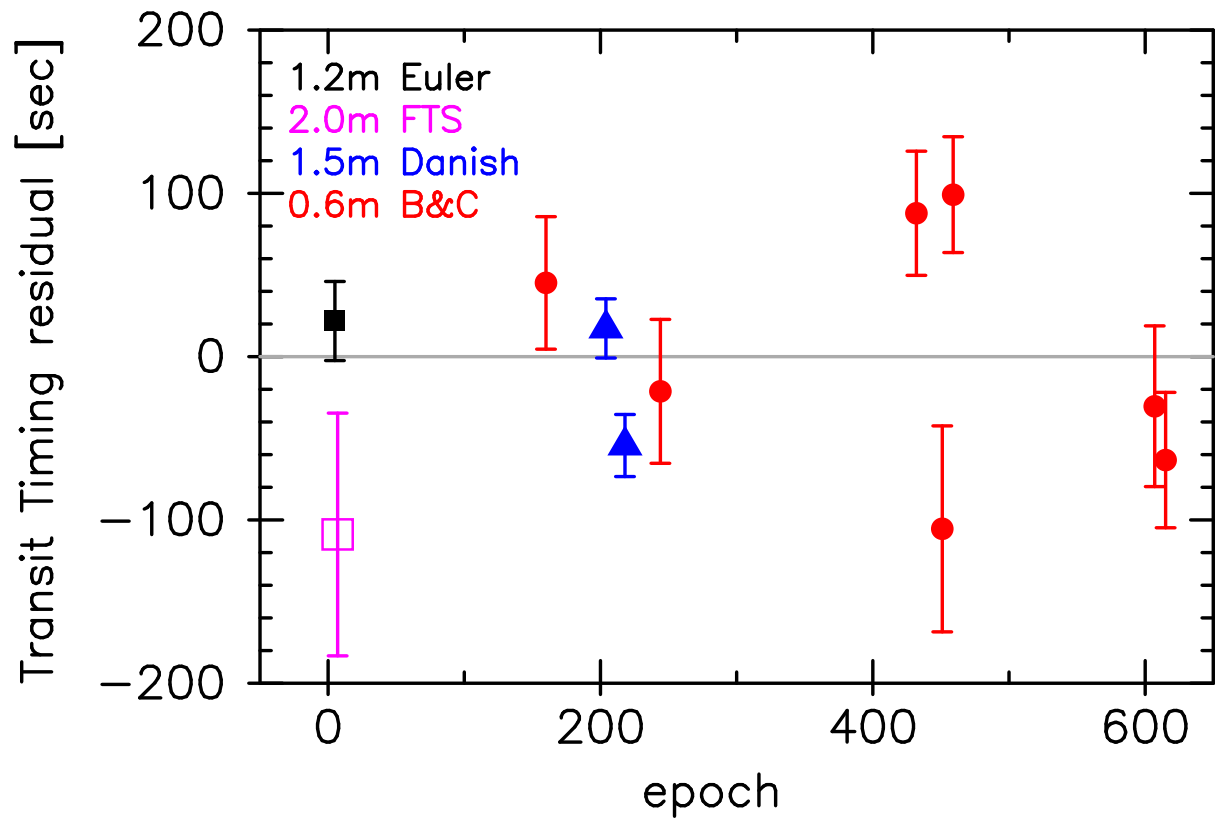}
\end{center}
\caption{
Transit Timing residuals of WASP-5b using a new ephemeris determined in section \ref{sec:TTV}. A filled square is the Euler data, an open square is  the FTS data, two triangles are the Danish data, and seven filled circles are the B\&C data.
\label{fig:oc}}
\end{figure}

We show the transit timings with 1-$\sigma$ uncertainties in table \ref{tbl:TTVs}. Using these data, a new ephemeris was calculated via a linear fit using a function
\begin{eqnarray}
T_{\mathrm{c},E} = T_{\mathrm{c},0} + EP,
\end{eqnarray}
where $T_{\mathrm{c}, 0}$ is a reference time of $E$=0.
The results are $T_{\mathrm{c}, 0}$ [BJD$_\mathrm{TDB}$] = 2454375.62510 $\pm$ 0.00019  and 
$P$ = 1.62843142 $\pm$ 0.00000064 d.
This fit yields $\chi^2 = 32.2 $ for 9 degrees of freedom which indicates that 
a linear fit does no mach at a confidence level of 99.982 \%, or 3.7 $\sigma$.
Figure \ref{fig:oc} shows the timing residuals and their uncertainties.
This result implies that we see excess variations in the 
transit times due either to unknown systematic effects or possibly to real TTVs.
%The possibility of existing a TTV signal in this system  was first implied by G09 as $\chi^2_\mathrm{/dof}$ for a linear fit of 5.7 with four timing data (corresponding to a significance of 2.9 $\sigma$), and 
A liner fit for only the B\&C data yields $\chi^2$ = 17.1 for 5 degrees of freedom which corresponds to a confidence level  of 99.57 \% or 2.9 $\sigma$, while the fit for the other 4 data yields $\chi^2$ = 9.8 which corresponds to a confidence level of 99.27 \% or 2.7 $\sigma$.
Even if two data having the largest uncertainties in all 11 timing data (corresponding epochs of  $E$=7 and 451), which  may have larger systematics, are discarded, the $\chi^2$ value for a linear fit remains 27.1 for 7 degrees of freedom, which corresponds to a confidence level of 99.968 \% or 3.6 $\sigma$.
This  result increases the statistical significance of a TTV excess implied by G09, who used four timing data (two of them were derived from same photometric data as we use) and derived $\chi^2_\mathrm{/dof}$ for a linear fit of 5.7 which corresponds to a significance of 99.67 \% or 2.9 $\sigma$.

The standard deviation of the observed timing residuals is 68 s and the mean uncertainty of the timings is 41 s, therefore if this is a real TTV signal, the actual standard deviation of the TTV is expected to be  $\sim$50 s. Such a large timing deviation could not be explained by effects other than additional perturbing bodies. The size of these alternative possibilities are: the Applegate effect \cite[$\sim1.5$ s per 11 years,][]{2010MNRAS.405.2037W}, the light travel time effect due to an outer massive body \cite[$\lesssim$ 1 s per 3 years, e.g.,][]{2005MNRAS.359..567A}, orbital decay caused by tidal dissipation \cite[$\sim$ 5 ms year$^{-1}$,][]{2009Natur.460.1098H}, and orbital precessions due to tidal deformations ($<$ 130 ms year$^{-1}$), due to the general relativity effect ($<$ 2 ms year$^{-1}$), and due to stellar quadrupole moment \cite[$<$ 0.005 ms year$^{-1}$,][]{2007MNRAS.377.1511H,2008ApJ...685..543J} (see table \ref{tbl:other_effects}). 
In addition, the presence of an exomoon also could not be responsible for the signal, because \citet{2010arXiv1007.4500W} placed an upper mass and distance (9.1$\times$10$^{-4}$ M$_{\oplus}$ and 4.39 R$_\mathrm{Jup}$ respectively) of a possible moon for the WASP-5b system by considering the three-body stability issue; such moon would result in only $\sim$0.02 s TTVs.
Thus, the remaining possibilities for explaining the signal are additional planets or Trojan companions (bodies at 1:1 MMR).

\subsection{Upper Mass Limits for Secondary Planet}
A marginal evidence for TTV signal is seen, however,  it is hard to determine the characteristics of the additional body (planet) at this time because of a large parameter space for the planet parameters: mass, orbital period, eccentricity, position of periapsis, reference phase, and mutual inclination. Alternatively, using the observed TTV data, we place upper limits on the mass of the hypothetical additional planet as a function of period ratio of the two planets (WASP-5b and the secondary body) by 
numerical thee-body simulations,
similar to several previous analyses  \cite[e.g.,][]{2005MNRAS.364L..96S, 2008ApJ...682..586M, 2008ApJ...682..593M, 2009A&A...506..369B, 2009ApJ...700.1078G, 2010ApJ...714...13A, 2010MNRAS.403.2111H, 2010MNRAS.401.1917G}.

In order to simplify things, we assume that the secondary planet orbits in the same plane as WASP-5b (the primary planet) and that the two planets have initially circler orbits. 
Because the amplitude of timing variations tends to increase as the orbital eccentricity of the second planet increases, basically we only need to consider an initially circular orbit for the secondary planet \cite[e.g.,][]{2010MNRAS.401.1917G}. 
For a given initial orbital period, mass ($M_\mathrm{2}$), and initial phase ($\theta_\mathrm{2}$) of the secondary planet,
we perform numerical integrations of the equations of motion for the three-body system using the 4th-order Runge-Kutta method and a time step of 1 s. We continue the integrations for an equivalent elapsed time of 3,000 days, which is three times longer than the observation time interval.
When the primary planet in the numerical procedure passes through the star-observer field of view the time step is reduced to 0.001 s and the calculations are used to create simulated transit timing data.
These data are fitted to a linear function in order to obtain an ``observed" orbital period and synthetic TTV data, which are then compared to the observed TTV data and a $\chi^2$ value obtained.
Transit timings of the secondary planet are also extracted in order to calculate its observed orbital period.

For a given secondary-planet model (initial period ratio, mass, and phase),  a $\chi^2$ value is chosen so as to be a minimum by shifting the reference epoch of the artificial TTV data sequentially.
We also assume that the TTV amplitude is proportional to $M_\mathrm{2}$ \citep{2005MNRAS.359..567A, 2005Sci...307.1288H} and integrate only the case of $M_\mathrm{2} = 3 \times 10^{-6} M_{\odot}$ for each initial period ratio (ranging from 0.2 to 5.0, increased by a factor 1.005)  and $\theta_\mathrm{2}$ (stepped by 30 degree), then rescale $M_\mathrm{2}$ (or the amplitudes of the  artificial TTV data) so that the  $\Delta \chi^2$, between the $\chi^2$ value derived above and a $\chi^2$ value for linear fit,  becomes 9.0 (3 $\sigma$ confidence limits).

To check the long term stability, we also conduct long term runs for up to 10$^5$ years with a time step of 100 s and eliminate models in the cases the system becomes unstable (one body is ejected or two bodies collide).
We then derive 12 upper-limit values of M$_\mathrm{2}$  from 12 models of $\theta_\mathrm{2}$ as a function of observed period ratio, and  derive a maximum upper-limit value by taking a maximum value for a given period ratio from the linearly-interpolated upper-limit lines.

\begin{figure}[ht]
\begin{center}
\FigureFile(160mm,100mm){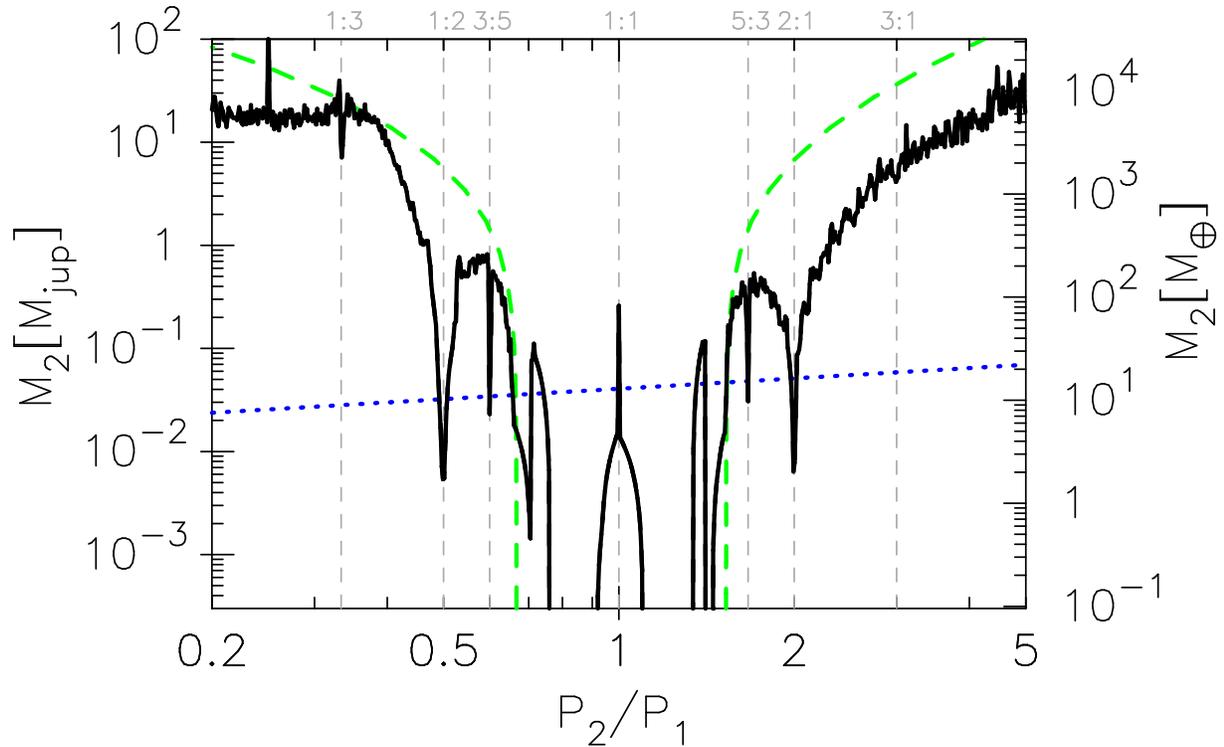}
\end{center}
\caption{
Estimated 3 $\sigma$ upper limits on mass of the hypothetical secondary planet as a function of orbital-period ratio (black thick-solid line), assuming two planets are co-planer and have initially circular orbits.
The dotted line represents upper limits on the secondary mass estimated from the RV data, which corresponds to a line causing the RV amplitude of  21~m~s$^{-1}$ due to the secondary body.
The thick-dashed line shows boundary of Hill-stable region calculated from \citet{2006ApJ...647L.163B}. 
Gray vertical-dashed lines represent corresponding MMRs.
\label{fig:limit}}
\end{figure}

The derived 3 $\sigma$ upper limits are shown in figure \ref{fig:limit} (black solid line).
The overplotted thick-dashed line represents the boundary of Hill-stable regions calculated from \citet{2006ApJ...647L.163B}, inside which long-term stabilities are not guaranteed while some stable regions can still exist \citep{2006ApJ...647L.163B}.
We also overplot an upper-limit line estimated from the RV data (dotted line), which corresponds to a line causing the RV amplitude of 21~m~s$^{-1}$; we assume that the added systematic error of 7~m~s$^{-1}$ is entirely due to the possible additional body and multiply this value by three as an upper limit.
As a result, outside the boundaries, we place upper limits on the secondary mass as 22-70 M$_{\oplus}$ in the period ratio from 0.2 to 5.0 from the RV data, and more stringent limits down to 2 M$_{\oplus}$ near 1:2 and 2:1 MMRs from the TTV data (at the 3 $\sigma$ level respectively).

We also put the upper limit on the mass of a Trojan companion (an object at one of the two triangular Lagrange points, L4 and L5,  of the planet-star orbit) as 100 M$_{\oplus}$ (3 $\sigma$) from the TTV data. 
The upper limit of $\sim$40 M$_{\oplus}$ at 1:1 MMR from the RV data is not correct, because if a Trojan companion exists at L4 or L5 points, the observed RV data would show one planet orbiting at the barycenter of the primary planet and the Trojan companion. 
Instead, a method to find an imbalance of mass at the L4/L5 points using both RV and photometric data has been proposed by \citet{2006ApJ...652L.137F}.
The basic idea of this method is to observe a difference between the time of vanishing stellar RV variation ($T_\mathrm{0, RV}$) and the time of the mid-transit ($T_\mathrm{c}$). \citet{2009ApJ...693..784M} estimated the upper mass of the imbalance mass of Trojan companions in the WASP-5 system as $<$54.7 M$_{\oplus}$ (2 $\sigma$) at the L5 point (behind the planet), based on the data of A08. Here we improve this estimate by employing more RV and photometric data. We measure the time difference $\Delta t$ = $T_\mathrm{c,}$$_0$ - $T_\mathrm{0, RV}$
with (without) RV data during transit (14 data points) adopting the eccentricity of WASP-5b is zero, resulting $\Delta t$ = 4.8 +/- 4.4 min (12.8 +/- 4.8 min).
Accordingly, we set upper limits on the excess mass of the Trojan companions, $M_\mathrm{T}$, which is defined as the difference in the mass at L4 ($M_\mathrm{T, L4}$) and the mass at L5 ($M_\mathrm{T, L5}$) (namely, $M_\mathrm{T} \equiv M_\mathrm{T, L4} - M_\mathrm{T, L5}$), through the relation;
\begin{eqnarray}
M_\mathrm{T} = M_\mathrm{p} \left( \frac{2 \tan(2 \pi \Delta t / P)}{ \sqrt{3} - | \tan(2 \pi \Delta t /P)|} \right),
\end{eqnarray}
where $P$ and $M_\mathrm{p}$ are the orbital period and mass of WASP-5b respectively [ equation (2) in \cite{2009ApJ...693..784M} and originally from equation (1) in \cite{2006ApJ...652L.137F} ]. We find $M_\mathrm{T}$ = 7.4 $\pm$ 6.8 (19.8 $\pm$ 7.7) $M_{\oplus}$ and put an upper limit on the excess mass  near the L4 point as $\sim$28 $M_{\oplus}$ ($\sim$43 $M_{\oplus}$) at the 3 $\sigma$ level. This result lowers the limit derived from only the TTV data ($<$ 100 $M_{\oplus}$), and improve the one derived by \citet{2009ApJ...693..784M} by at least a factor 2.

\subsection{Example Models of Secondary Planet}
\begin{figure}[ht]
\begin{center}
\FigureFile(83mm,100mm){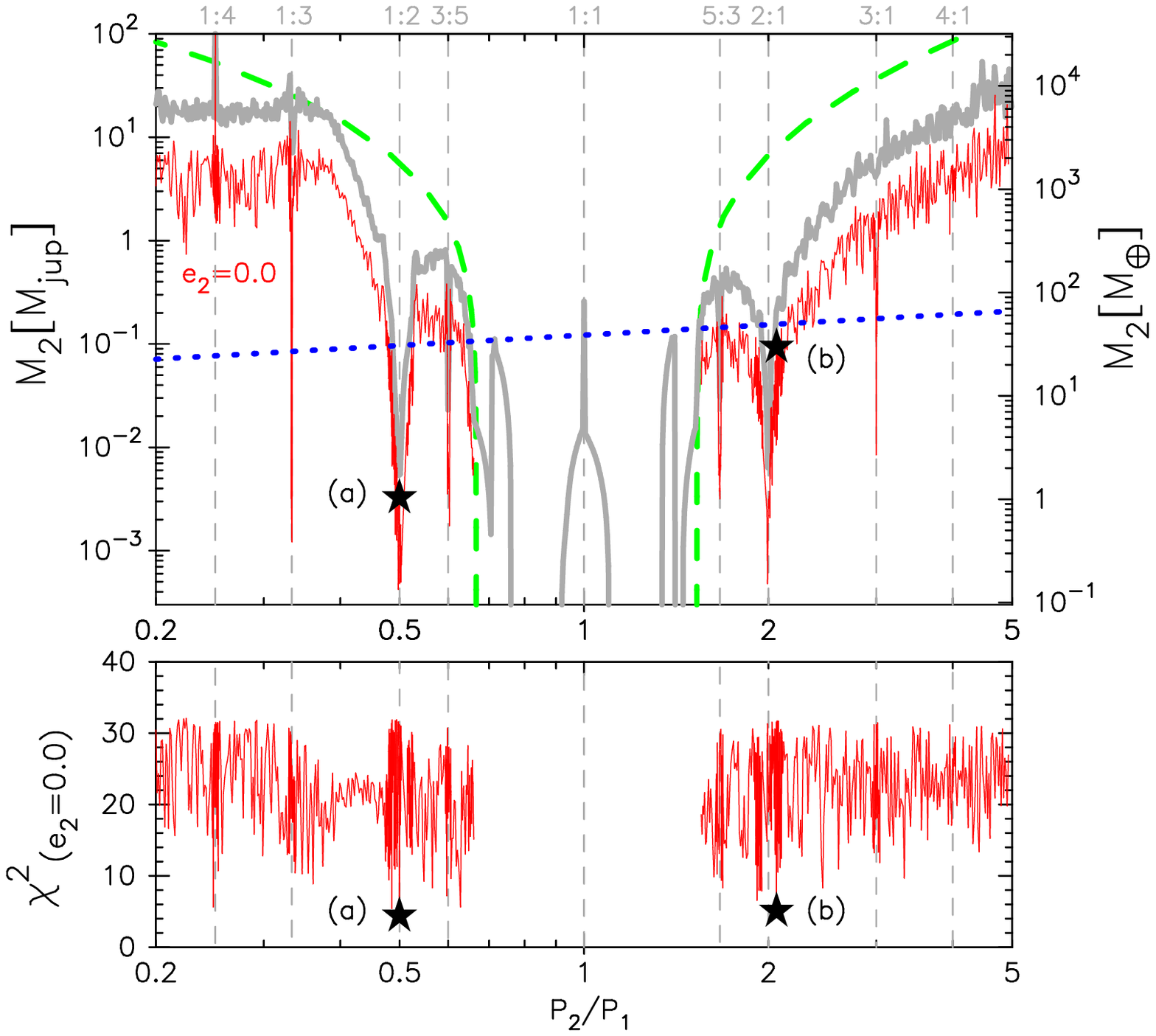}
\FigureFile(83mm,100mm){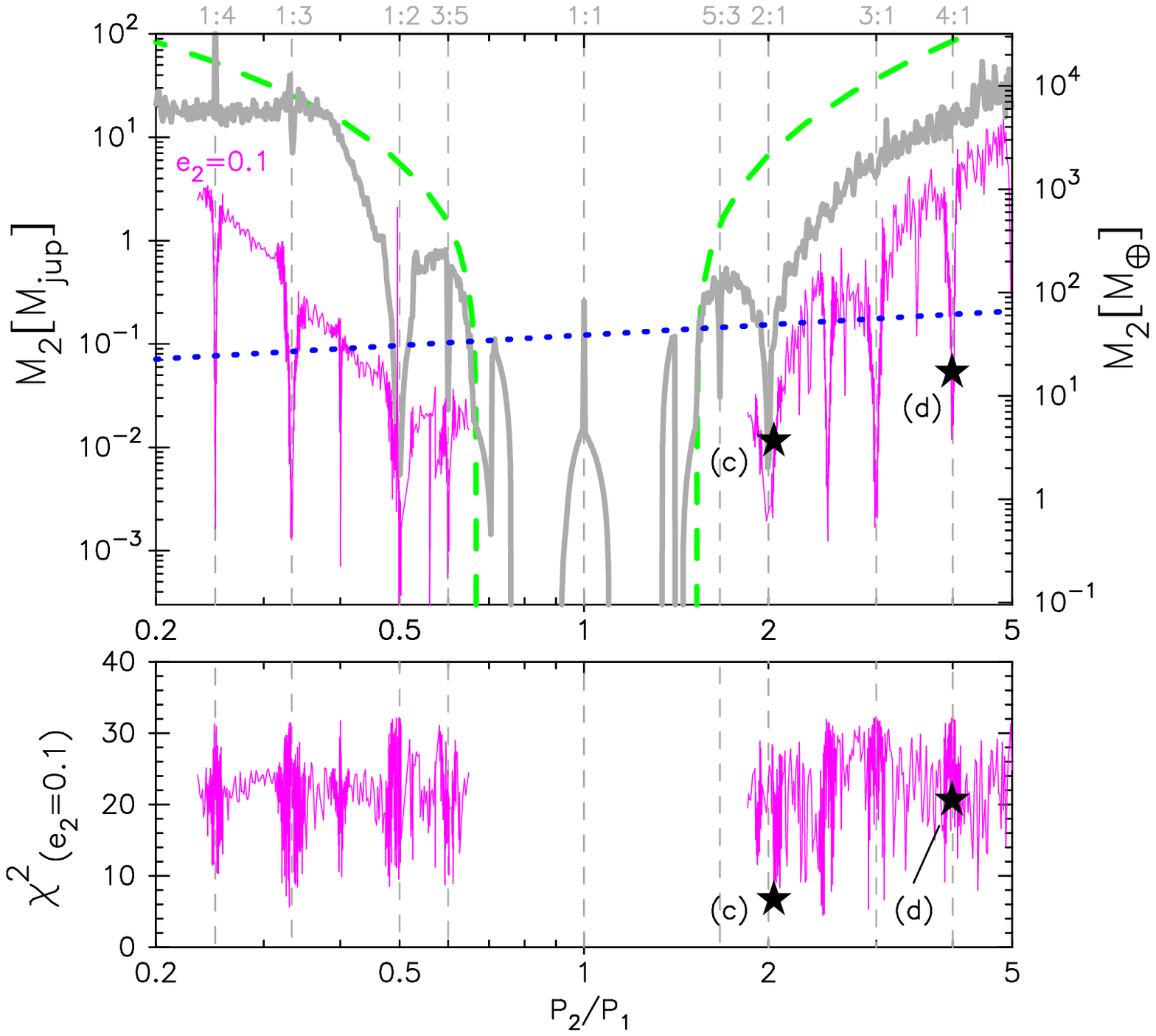}
\end{center}
\caption{
The best-fit mass regions of the hypothetical secondary planet as a function of orbital period ratio (thin-solid lines in upper panels) and their $\chi^2$ values (lower panels), for the case of $e_2$ = 0 (left) and 0.1 (right). The gray thick lines are same as the black thick line in figure \ref{fig:limit}. The locations of four example TTV models shown in figure \ref{fig:bestfit_solutions} are marked as star symbols and alphabets of (a), (b), (c), and (d).
\label{fig:bestfit_regions}}
\end{figure}

Here, in order to illustrate example models of secondary planet which can account for the observed timing variations,  we fit the simulated TTV data to the observed one and search for well-fitting models.
%, although a number of solutions for the secondary-planet parameters are degenerated under the low significant TTV signal. 
Using the artificial TTV data generated in the previous section,
 in which the initial eccentricity of the secondary planet, $e_2$, was set to zero, 
 we search for a best-fit solution for each orbital period ratio by scaling TTV amplitude and shifting a reference epoch of the artificial TTV data. 
At this time, we use only the case that the initial phase of the secondary planet is zero for simplicity. 
 Since, generally, a libration period of TTVs increases as the orbital period ratio approaches to a MMR, solutions around MMRs should be searched finely. For this reason, we generate a set of additional artificial TTV data around the low-order MMRs of 1:4, 1:3, 1:2, 3:5, 5:3, 2:1, 3:1, and 4:1, by 10 times denser than the other regions, and search for best-fit solutions also for the additional period ratios.
 We then convert the best-fit TTV amplitudes to the secondary masses, 
assuming a TTV amplitude is proportional to a secondary-planet mass.
In addition to the case of $e_2$=0, we also generate a set of artificial TTV data for the case of $e_2$=0.1 in the same conditions as the previous one and in an additional condition; initial phase of the periapsis of the secondary planet which is set to zero. In this case, the regions around the MMRs of 1:4, 1:3, 2:5, 1:2, 3:5, 2:1, 5:2, 3:1, and 4:1 are searched densely.
We then search for the best-fit solutions against the respective orbital period ratios. 
Figure \ref{fig:bestfit_regions} shows the derived best-fit mass of the secondary planet as a function of the orbital period ratio for the cases of $e_2$=0 (left top panel) and $e_2$=0.1 (right top panel), and their $\chi^2$ values (bottom panels). 

\begin{figure}[ht]
\begin{center}
\FigureFile(83mm,100mm){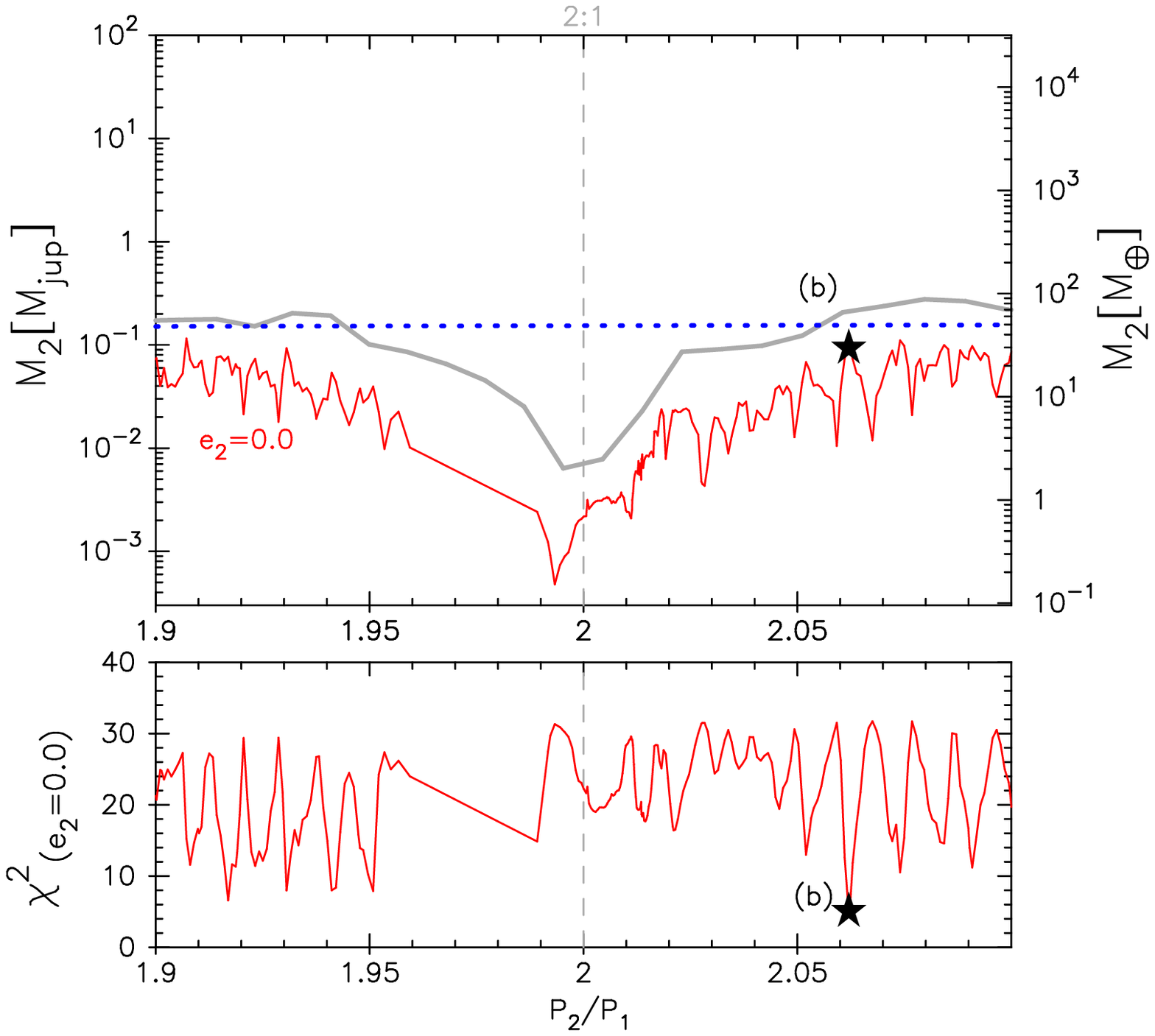}
\FigureFile(83mm,100mm){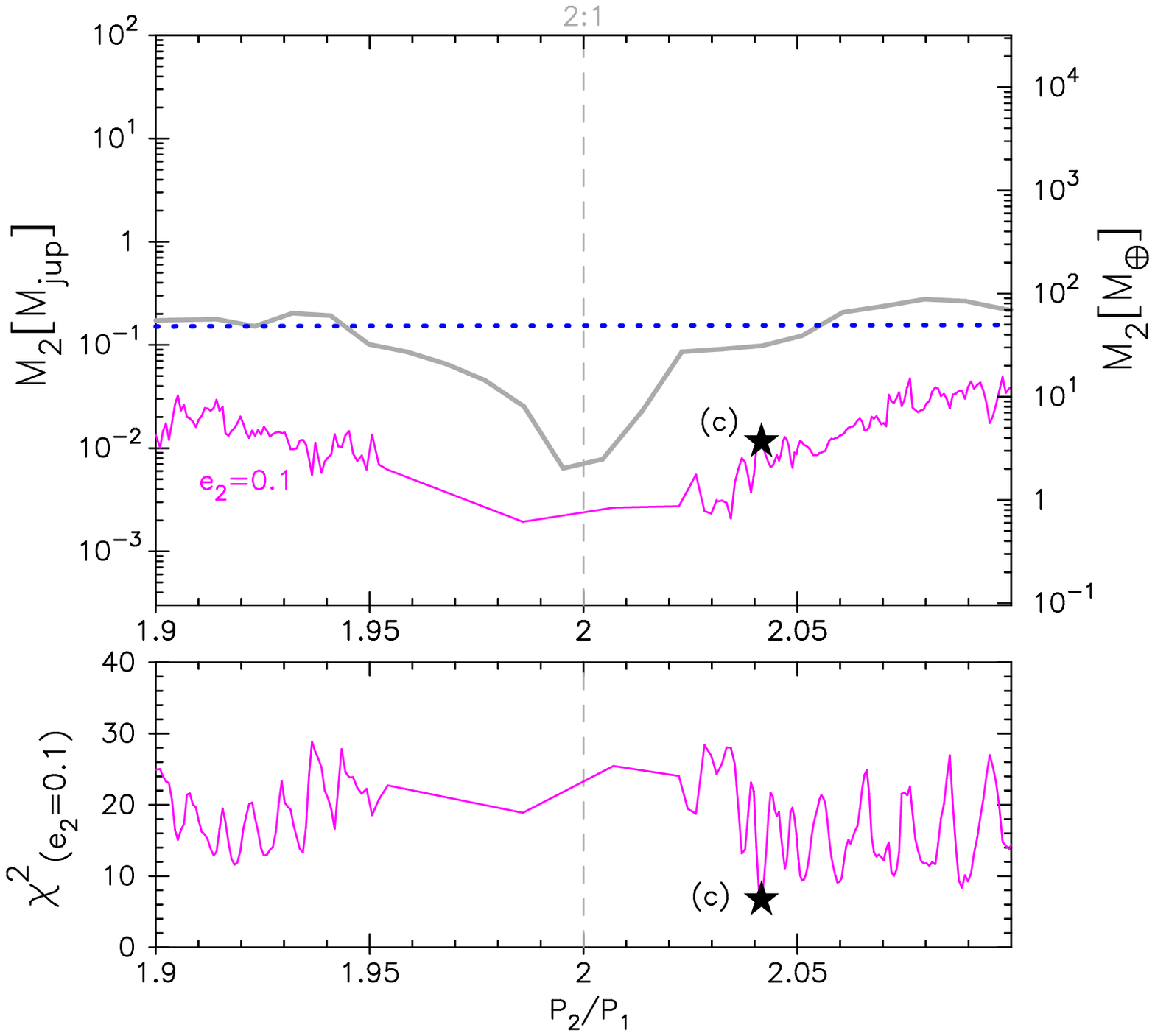}
\end{center}
\caption{Zooms around the 2:1 MMR of figure \ref{fig:bestfit_regions}.
The pitches of the ``observed'' period ratio are enlarged in the vicinity of the 2:1 MMR, because the strong resonance much excites the orbital eccentricities of the two planets resulting in the jumps at a certain period-ratio range.
\label{fig:zoom_MMR}}
\end{figure}

As a result, for the case of $e_2$=0,
the regions where the best-fit masses are lower than the upper limits placed from the radial velocity data (RV limits) are limited only around the low-order MMRs of 1:3, 1:2, 3:5, 5:3, 2:1, and 3:1. Such regions then extend to around other MMRs as $e_2$ increases to 0.1. 
%The $\chi^2$ maps of the best-fit models are shown in the bottom panels of figure \ref{fig:bestfit_regions}. 
In addition, the $\chi^2$ maps show there are a number of local minima over the period-ratio range,
%appear jagged shapes across the horizontal axis, 
reflecting the fact that the libration period of artificial TTVs gradually changes as the orbital period ratio changes, and some of the local minima around MMRs have low-$\chi^2$ values, less than the degrees of freedom of 9.
Figure \ref{fig:zoom_MMR} shows zoom around the 2:1 MMR of figure \ref{fig:bestfit_regions} for example.
These facts suggest that it is possible to explain the observed TTV data by a certain model of a perturbing planet having lower mass than the RV limit,
and such a model would likely be near low-order MMRs, if its eccentricity is enough small.
Non- or a small eccentricity of the secondary planet is naturally expected if the planet has migrated by the disk-planet interactions, 
and if the above scenario is true, this fact  could be important observational evidence that the hot Jupiter has been formed according to one of the disk-planet interaction models \citep[e.g.,][]{2005ApJ...626.1033T}.
This would also be consistent with the fact that the sky-projected spin-orbital angle is consistent with zero.

\begin{figure}[ht]
\begin{center}
\FigureFile(75mm,60mm){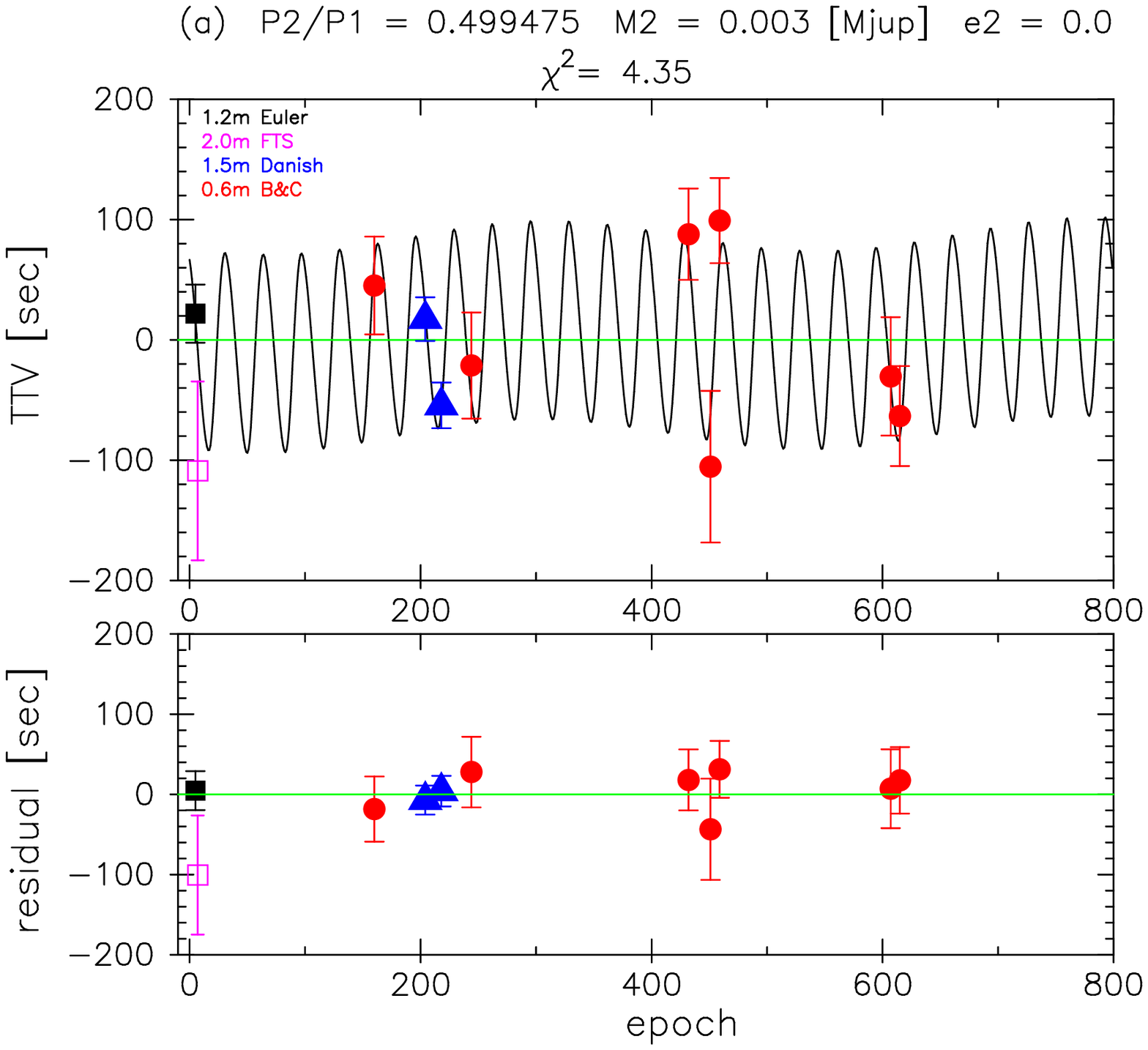}
\FigureFile(75mm,60mm){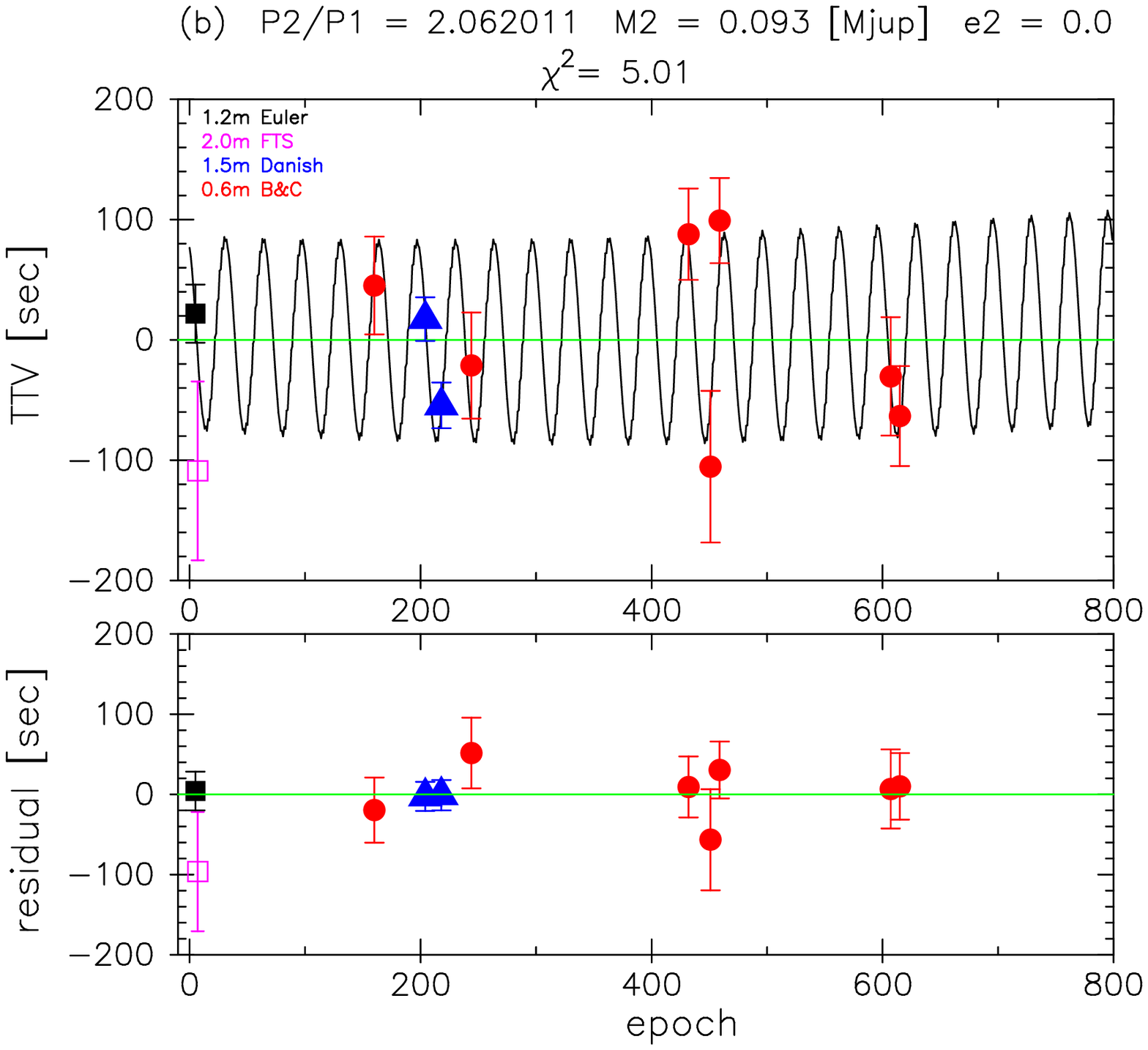}\\
\FigureFile(75mm,60mm){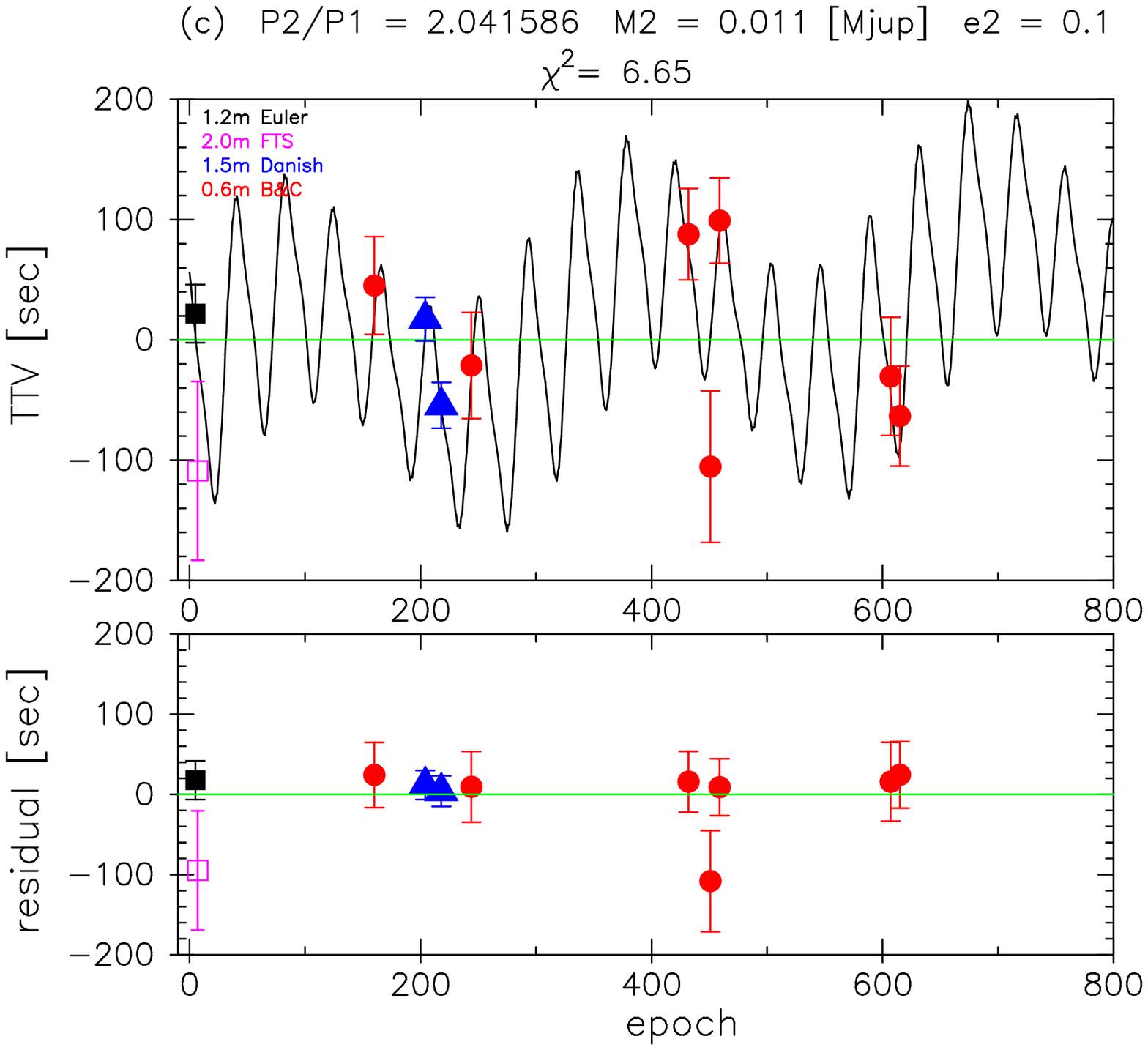}
\FigureFile(75mm,60mm){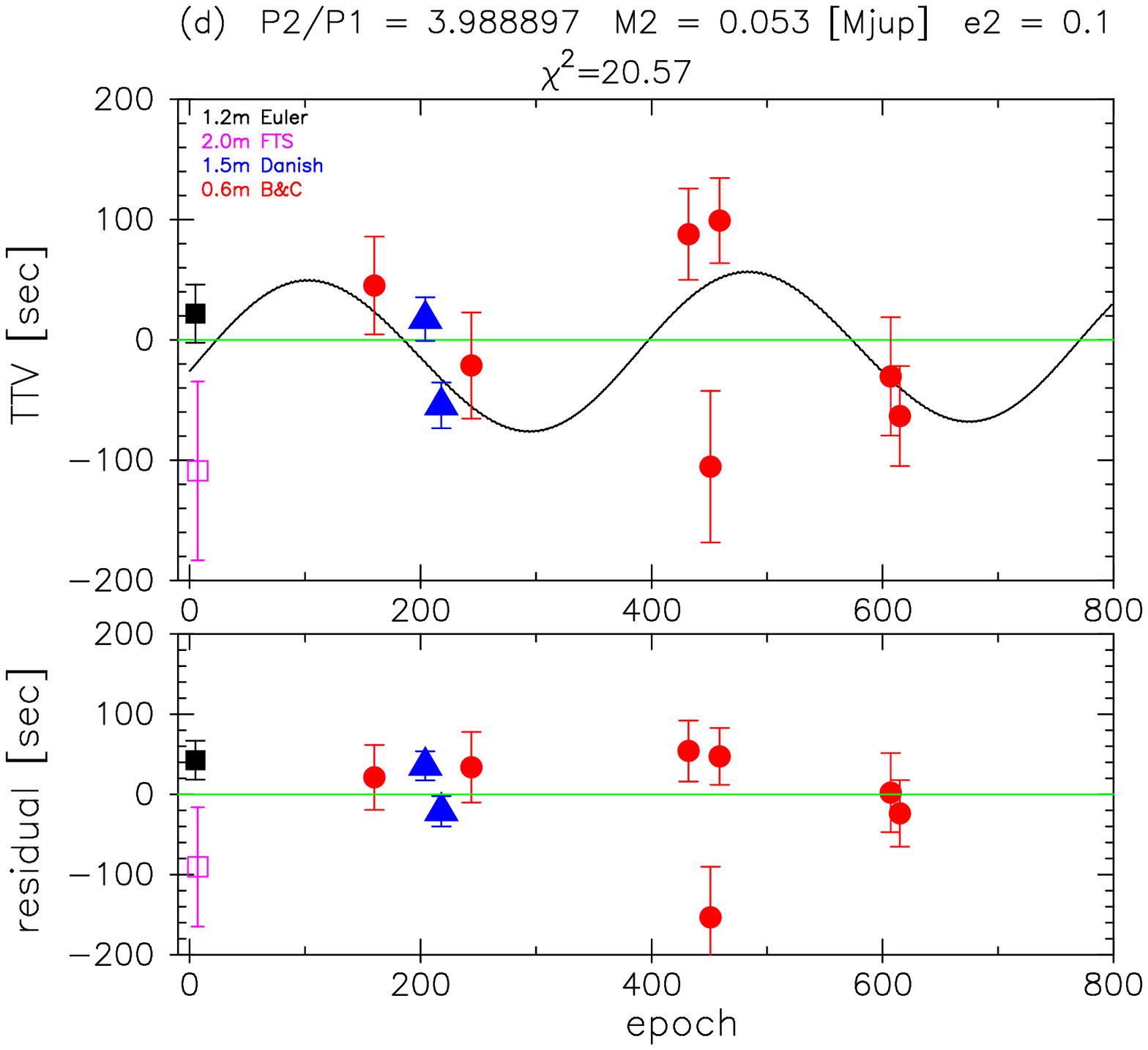}
\end{center}
\caption{
Four example TTV models (black curves) with the observed TTV data (legends are same as figure \ref{fig:oc}) (top panels) and their residuals (bottom panels).  
The explanations for these models appear in the text.
%(a) is a model of $e_2$=0, $P_2/P_1$=0.499 (near 1:2 MMR), and $M_2$=0.003$M_\mathrm{jup}$. The $\chi^2$ value is 4.35 which is the minimum in all models considered here. (b) is a model of $e_2$=0, $P_2/P_1$=2.06 (near 2:1 MMR), and $M_2$=0.093$M_\mathrm{jup}$, with the $\chi^2$ value of 5.01. (c)  is a model of $e_2$=0.1, $P_2/P_1$=2.04 (near 2:1 MMR), and $M_2$=0.011$M_\mathrm{jup}$, with the $\chi^2$ value of 6.65.
%(d)  is a model of $e_2$=0.1, $P_2/P_1$=3.99 (near 4:1 MMR), and $M_2$=0.0053$M_\mathrm{jup}$, with the $\chi^2$ value of 20.6. When two timing data which have the largest error bars are discarded, the $\chi^2$ value reduces to 11.6. The masses of these four models are all lower than the upper limits placed from the RV data. 
}
\label{fig:bestfit_solutions}
\end{figure}

We show four example TTV models (a, b, c, and d), having lower masses than the RV limits and locating near MMRs, with the observed TTVs in figure \ref{fig:bestfit_solutions}.  The locations in parameter space of period ratio and mass are marked as star symbols in figure \ref{fig:bestfit_regions}.
(a) is the least $\chi^2$ model ($\chi^2$=4.35) in all models considered here, which is  located near the 1:2 MMR and 0.003 $M_\mathrm{jup}$ in the case of $e_2$=0. (b) is a low-$\chi^2$ model ($\chi^2$=5.01) located near the 2:1 MMR  and 0.093 $M_\mathrm{jup}$ in the case of $e_2$=0. Since (a) and (b) show similar libration periods and amplitudes, distinguishing them by only ground-based TTV observations would be difficult and radial velocity followup observations would be necessary, if one of such models is true. (c) is another low-$\chi^2$ model ($\chi^2$=6.65) located near the 2:1 MMR and 0.011 $M_\mathrm{jup}$ in the case of $e_2$=0.1. (d) 
is a model located near the 4:1 MMR and 0.053 $M_\mathrm{jup}$ in the case of $e_2$=0.1 with $\chi^2$=20.57. The $\chi^2$ value of the model (d) is somewhat high due to the relatively longer libration period, however, if two timing data having the largest error bars ($E$=7 and 451) are discarded, the $\chi^2$ value decrease to 11.6 for 7 degrees of freedom, and therefore such longer libration-period models are also thinkable.

On the other hand, non-MMR with rather higher eccentricity models might also be possible for which we do not search here.
If this case is true, it would be the first example where a hot-Jupiter system has a non-MMR, low-mass planet with a short orbit.
 Recently it has been revealed that most multi-planetary systems consisting of several hot super-Earths are not in MMRs \citep{2009A&A...507..487M, 2009A&A...493..639M} and this can not  be explained naturally using standard migration models \citep{2007ApJ...654.1110T, 2010arXiv1008.0461O}. Thus, it is important to check whether the situation in hot-Jupiter systems is same or not.

%   In figure \ref{fig:bestfit_solutions}, we illustrate four example TTV models  with the observed TTV data and their residuals.  The first one (a) is the lowest mass model in the nine least-$\chi^2$ models for $e_2$=0 which locates at a period ratio of 2.06 (near 2:1 MMR) and mass of 0.059$M_\mathrm{jup}$, which is a bit above the RV limit. The second and third ones (b, c) are the two lowest-mass ($M_\mathrm{jup}$ and $M_\mathrm{jup}$) models in the eight least-$\chi^2$ models for $e_2$=0.1, which locate at period ratios of 0.59 (near 5:3 MMR) and 0.50 (1:2 MMR), respectively. The final one (d) is a model of $e_2$=0.1 which locates at  a period ratio of 3.99 (near 4:1 MMR) and mass of $M_\mathrm{jup}$, just bellow the RV limit. This model has a $\chi^2$ value of  and is not  one of the least-$\chi^2$ models, but a model showing a long periodicity of TTVs
% These four models are all lie near low-order MMRs, and this fact implies that the (near-) MMR models would be favored rather than non-MMR models for such low-eccentricity cases.
% If this is true, the fact  could be important observational evidence that the hot Jupiter has been formed according to one of the disk-planet interaction models \citep[e.g.,][]{2005ApJ...626.1033T}.
%This would also be consistent with the fact that the sky-projected spin-orbital angle is consistent with zero. 

In order to distinguish these models and find an exclusive solution, additional high-precision and extensive timing data are necessary, and high-precision radial velocity followup observations are also helpful to confirm the additional planetary signal and constrain the physical parameters.
A successful program would provide valuable information for improving our understandings of planetary formation scenarios.

%---------------------------------------------
\section{Summary}
\label{sec:conclusion}
We have observed  7 new transits of the hot Jupiter WASP-5b using  61 cm telescope in order to search for an additional body via the TTVs method.
By combining all available photometric and RV data, we slightly refined the transit-model parameters and confirmed all parameters were consistent with published values, except for the orbital period which may be a sign of TTVs. 
We also confirmed that the sky-projected spin-orbit angle is consistent with zero, which was first reported by \cite{2010arXiv1008.2353T}, by including independent photometric data and using an improved RM formula. This fact is consistent with a disk-planet interaction model being the favored migration scenario for  WASP-5b, although planet-planet scattering models cannot be excluded at this point in time.

The $\chi^2$ value for a linear fit to the 11 transit timings is 32.2 for 9 degrees of freedom. This result indicates that the transit timings do not match a linear ephemeris at a confidence level of 99.982 \%, or 3.7 $\sigma$, due either to unknown systematic effects or possibly to real TTVs.
If this signal is real, the standard deviation of the TTVs is as large as 50 s, and the most likely cause of this is the  presence of an additional body in the system.
From the RV data, we put an upper limit on the RV amplitude caused by the possible secondary body (planet) as 21~m~s$^{-1}$, which corresponds to its mass of  22--70 M$_{\oplus}$ over the orbital period ratio of the two planets from 0.2 to 5.0.
From the TTV data, using the numerical simulations, we place more stringent limits down to 2 M$_{\oplus}$ near the 1:2 and 2:1 MMRs with WASP-5b at the 3 $\sigma$ level, assuming the two planets are co-planer. 
We also put the upper limit on excess Trojan mass as 43 M$_{\oplus}$ (3 $\sigma$) using both RV and photometric data.

We also find that if the possible secondary planet has non- or a small eccentricity, it would likely be near low-order MMRs. The fact would also support the disk-planet interaction models as the migration mechanisms for the hot Jupiter.
Alternatively, if the secondary body is an eccentric, non-MMR planet, it would be a challenge for the current planetary migration models.
Further followup observations for the WASP-5 system by both photometry and spectroscopy will reveal  the true nature of the TTV signal and shed light on the migration mechanisms of planetary systems.

%---------------------------------------
%\acknowledgments
\bigskip
We thank Alan Gilmore for general support at the Mt.\ John University Observatory. We acknowledge the University of Canterbury for allowing MOA to use the B\&C telescope.  We also acknowledge D.~R.~Anderson for providing the FTS and Euler data, and   M.~Gillon, J.~Southworth, and A.~H.~M.~J.~Triaud for providing information about the time systems in their published data.  This work was supported by the Global COE Program of Nagoya University ``Quest for Fundamental Principles in the Universe (QFPU)" from JSPS and MEXT of Japan. This work was partially supported by NAOJ Visiting Scholar Program (Short-term). 
A.F., N.N. and T.H. are supported by a Japan Society for Promotion of Science (JSPS) Fellowship for Research (DC1: 20-56341, PD: 20-8141, DC1: 22-5935).
This work is supported by the grant JSPS20340052. 
MT is supported by the Ministry of Education, Science, 
Sports and Culture, Grant-in-Aid for Specially Promoted Research, 22000005. 

%---------------------------------------
\appendix
\section*{}

When we consider primary and secondary atmospheric extinctions,
time-averaged standard magnitudes  of a target star and a reference star in an arbitrary passband,  $M_\mathrm{obj}$ and $M_\mathrm{ref}$, can be expressed as
\begin{eqnarray}
M_\mathrm{obj} &=& m_\mathrm{obj} - k\timeform{'} z - k\timeform{"} C_\mathrm{obj}  z + T C_\mathrm{obj} +  Z_\mathrm{p} +  f_\mathrm{obj}(t),\\
M_\mathrm{ref} &=& m_\mathrm{ref} - k\timeform{'} z - k\timeform{"} C_\mathrm{ref}  z + T C_\mathrm{ref} + Z_\mathrm{p} +  f_\mathrm{ref}(t),
\end{eqnarray}
where $m_\mathrm{obj}$ and $m_\mathrm{ref}$ are observed magnitudes of the target star and the reference star, $C_\mathrm{obj}$ and $C_\mathrm{ref}$ are their color indices, $z$ is the airmass, $T$ is the transformation factor,  Z$_\mathrm{p}$ is nightly zero point, $k\timeform{'}$ and $k\timeform{"}$ are coefficients of primary and secondary extinctions, and $f_\mathrm{obs}(t)$ and $f_\mathrm{ref}(t)$  are terms of stellar intrinsic variability as a function of time \citep[see e.g., Chapter 6 of][]{2006pglp.book.....W}. 
When we approximate that each stellar intrinsic brightness varies with time in linear function,
then the differential magnitude between the target star and the reference star, $\Delta M = M_\mathrm{obs} - M_\mathrm{ref}$, is written as
\begin{eqnarray}
\Delta M
                = \Delta m - k\timeform{"} \Delta C z  + T \Delta C + a t + b,
\end{eqnarray}
where $\Delta m = m_\mathrm{obj} - m_\mathrm{ref}$, $\Delta C = C_\mathrm{obj} - C_\mathrm{ref}$, and $a$ and $b$ are coefficients. 
Here, we assume that the $k\timeform{"}$ varies with time $t$ in linear function, as $k\timeform{"} = k\timeform{"}_\mathrm{0} + k \timeform{"}_\mathrm{1} t$, then
\begin{eqnarray}
\Delta M
                = \Delta m - (k\timeform{"}_\mathrm{0} + k\timeform{"}_\mathrm{1} t ) \Delta C z  + T \Delta C +  a t + b.
\end{eqnarray}
Consequently, if we redefine $k_\mathrm{z} \equiv - k\timeform{"}_\mathrm{0} \Delta C$, $k_\mathrm{tz} \equiv - k\timeform{"}_\mathrm{1} \Delta C$, $k_\mathrm{t} \equiv a$, and $k_\mathrm{0} \equiv T \Delta C + b$, then the correction magnitude $\Delta m_\mathrm{corr} = \Delta M - \Delta m$ can be written as
\begin{eqnarray}
                \Delta m_\mathrm{corr} = k_\mathrm{z} z + k_\mathrm{tz} t z + k_\mathrm{t} t + k_\mathrm{0}.
\end{eqnarray}

%---------------------------------------

%\bibliographystyle{hapj}
%	\bibliography{ref}

%---------------------------------------------------
%% tables

%% Observation Log
\begin{table}[ht]
\begin{center}
\caption{The observing Log}
\label{tbl:obslog}
\begin{tabular}{lcccccl}
\\
\hline
Date & epoch ($E$) & exp. [sec] &  filter & \# of data & focus/defocus & airmass\\
\hline
2008 June 18 & 160 & 60 & $I$ & 248 & focus & 1.55 $\rightarrow$ 1.00\\
2008 November 2& 244 & 30 & $I$ & 393 & focus & 1.00 $\rightarrow$ 1.35\\
2009 September 4 & 432 & 60 & $I$ & 168 & focus & 1.03 $\rightarrow$ 1.00 $\rightarrow$ 1.07 \\
2009 October 5 & 451 & 60 & $I$ & 313 & focus & 1.07 $\rightarrow$ 1.00 $\rightarrow$ 1.39\\
2009 October 18 & 459 & 60 & $I$ & 248 & defocus & 1.01 $\rightarrow$ 1.80\\
2010 June 16 & 607 & 30 & $I$ & 307 & focus & 2.60 $\rightarrow$ 1.23\\
2010 June 29 & 615 & 60 & $I$ & 202 & focus & 1.78 $\rightarrow$ 1.04\\
\hline
\\
\end{tabular}
\end{center}
\end{table}

%% OOT scatters
\begin{table}[ht]
\caption{Comparisons of rms scatters between OOT light curves derived using optimized radii ($\gamma_E r_{opt, j})$ and that derived using fixed radii ($r_{fix, E}$). $N_{OOT}$ is the number of OOT (either before or after the transit) data points and "b" or "a" in parentheses stands for before or after the transit, respectively.}
\begin{center}
\label{tbl:scatters}
\begin{tabular}{ccccccc}
\\
\hline
$E$ & $N_{OOT}$ &  \ \ \ \ $\gamma_E$ \ \ \ \ & $\gamma_E r_{opt, j}$ & \ \ \ \ $r_{fix, E}$ \ \ \ \ & \multicolumn{2}{c}{rms scatter for OOT (\%)}\\
&  & & (pixels) & (pixels) & $\gamma_E r_{opt, j}$ & $r_{fix, E}$\\
\hline
160 & 75 (a) & 0.96 & 9.6 - 13.1 & 13.1 & 0.177 & 0.178\\
244 & 119 (a) & 1.27 & 13.1 - 18.5 & 15.8 & 0.488 & 0.492\\
432 & 50 (b)  & 1.34 & 11.5 - 14.5 & 14.1 & 0.205 & 0.205\\
451 & 77 (a) & 1.47 & 14.4 - 16.0 & 14.0 & 0.282 & 0.284\\
459 & 52 (b)  & 1.24 & 13.9 - 17.1 & 13.9 & 0.179 & 0.181\\
607 & 49 (a) & 1.42 & 12.5 - 14.3 & 13.0 & 0.287 & 0.288\\
615 & 44 (a)  & 1.30 & 11.1 - 13.7 & 12.7 & 0.161 & 0.158\\
\hline
\end{tabular}
\end{center}
\end{table}

%% beta values
\begin{table}[ht]
\caption{Estimated red-noise factor $\beta$, the number of data points in one bin ($N_{max}$), un-binned rms scatter from the best-fit model ($\sigma_1$), and the total number of data points, for each transit light curve are listed. }
\begin{center}
\label{tbl:beta}
\begin{tabular}{cccccc}
\\
\hline
$E$ & $\beta$& $N_{max}$& $\sigma_1$ [\%] & \# of data & telescope\\
\hline
160  & 1.15	& 15	& 0.197	& 248	& B\&C\\
244  & 1.16	& 14 & 0.364	& 386	& B\&C\\
432  & 1.09	& 6   & 0.205	& 168	& B\&C\\
451  & 1.76	& 18	& 0.234	& 313	& B\&C\\
459  & 1.01	& 8	& 0.220	& 247	& B\&C\\
607  & 1.31	& 17	& 0.322	& 306	& B\&C\\
615  & 1.19	& 13	& 0.210	& 202	& B\&C\\
\\[-0.15in] 
5	& 1.16	& 10	& 0.094	& 138	& Euler\\
7	& 2.28	& 27	& 0.262	& 335 	& FTS\\
204	& 1.28	& 15	& 0.046	& 73		& Danish\\
218	& 1.34	& 15	& 0.054	& 101	& Danish\\
\hline
\\
\end{tabular}
\end{center}
\end{table}

%% limb-darkening coefficients
\begin{table}[ht]
\caption{The quadratic limb-darkening coefficients ($u_1$ and $u_2$) for each telescope used for the final joint fit are listed. Each $u_1$ was allowed to be free while each $u_2$ was fixed at the theoretical value extracted from tables of \citet{2000A&A...363.1081C} or \citet{2004A&A...428.1001C}.}
\label{tbl:limb-darkening}
\begin{center}
\begin{tabular}{cccc}
\\
\hline
Telescope	& filter	& $u_1$ (fitted)		& $u_2$ (fixed)\\
\hline
Euler	&	$R$ 		& 0.33 $^{+0.05}_{-0.05}$	& 0.32\\
FTS		& 	SDSS $i'$	& 0.49 $^{+0.12}_{-0.13}$	& 0.32\\
B\&C	&	$I$		& 0.27 $^{+0.05}_{-0.05}$	& 0.32\\
Danish	&	$R$		& 0.34 $^{+0.04}_{-0.04}$	& 0.32\\
\hline
\\
\end{tabular}
\end{center}
\end{table}

%% derived parametes
\begin{landscape} 
\begin{table}[ht]
\caption{The best-fit parameters and their 1-$\sigma$ uncertainties for WASP-5b derived from the joint fits for only the 7 light curves from the B\&C telescope (the second column) and for all available RV and photometric data (the third column) are listed. Parameters derived by S09 (the fourth column) and T10 (the rightmost column) are shown for comparison.
}
\begin{center}
\label{tbl:fit_params}
\begin{tabular}{ccccc}
\\
\hline
& This work (B\&C only) & This work (all) & \cite{2009MNRAS.396.1023S} & \citet{2010arXiv1008.2353T}\\
\hline
$P$ [days]			&  1.6284301 $\pm$ 0.0000012	& 1.62843142 $\pm$ 0.00000064	& 1.6284246 $\pm$ 0.0000013      & $1.6284229^{+0.0000044}_{-0.0000039}$\\
$T_0$ $[BJD_{TDB} - 2450000]$				& 4375.62589 $\pm$ 0.00052	& 4375.62510 $\pm$ 0.00019	& 4375.62569 $\pm$ 0.00024  & $4373.996764^{+0.00015}_{-0.00014}$\\
$i$ [$^{\circ}$]			& $85.01^{+1.17}_{-0.78}$		& $85.58^{+0.81}_{-0.76}$	& 85.8 $\pm$ 1.1	& $86.2^{+0.8}_{-1.7}$\\
$R_p/R_*$			& $0.1116^{+0.0010}_{-0.0014}$	& $0.1108 \pm 0.0011$	& 0.1110 $\pm$ 0.0014	& $0.1105^{+0.0019}_{-0.0007}$\\
$a/R_*$				& $5.26^{+0.23}_{-0.17}$			& $5.37 \pm 0.15$	& $5.41^{+0.17}_{-0.18}$	& $5.49^{+0.37}_{-0.12}$\\
$K$ [m s$^{-1}$]			& - & $269.4 \pm 3.3$ & - & $268.7^{+1.7}_{-1.9}$\\
$e$					& 0 (adopted) & 0 (adopted) & 0 (adopted) & $<$ 0.0371\\
$V \sin{I}$ [km s$^{-1}$]	& - & $3.05 \pm 0.41$ & - &	 $3.24^{+0.34}_{-0.35}$\\
$\lambda$ [$^{\circ}$]	& - & $7.2 \pm 9.5$ & - & $12.4^{+8.2}_{-11.9}$\\
\hline
\\
\end{tabular}
\end{center}
\end{table}
\end{landscape}

%% Transit Timings
\begin{table}[ht]
\begin{center}
\caption{Transit Timings and their uncertainties.}
\label{tbl:TTVs}
\begin{tabular}{cccc}
\\
\hline
E & Transit Timing  &1-$\sigma$ uncertainty & telescope\\
& [BJD$_{TDB}$ - 2 450 000] & \\
\hline
5	& 4383.76751	& 0.00028		& Euler\\
7	& 4387.02286	& 0.00086		& FTS\\
160	& 4636.17465	& 0.00047		& B\&C\\
204	& 4707.82531	& 0.00021		& Danish\\
218	& 4730.62252	& 0.00022		& Danish\\
244	& 4772.96212	& 0.00051		& B\&C\\
432	& 5079.10849	& 0.00044		& B\&C\\
451	& 5110.04645	& 0.00073		& B\&C\\
459	& 5123.07627	& 0.00041		& B\&C\\
607	& 5364.08262	& 0.00057		& B\&C\\
615	& 5377.10969	& 0.00048		& B\&C\\
\hline
\end{tabular}
\end{center}
\end{table}

%% physical parameters
\begin{table}[ht]
\caption{Derived physical values and their uncertainties for the WASP-5b system. The value of the stellar mass was adopted to the value derived by T10. The  values presented in S09 and T10 are shown for comparison. 
}
\begin{center}
\label{tbl:phys_params}
\begin{tabular}{cccc}
\\
\hline
& This work			& \citet{2009MNRAS.396.1023S}	& \citet{2010arXiv1008.2353T}\\
\hline
$a$ [AU]				& 0.02702 $\pm$ 0.00059	& 0.02729 $\pm$ 0.00049 $\pm$ 0.00027 & $0.02709^{+0.00056}_{-0.00062}$ \\
$R_*$ [R$_{\odot}$]		& 1.082 $\pm$ 0.038		& 1.084 $\pm$ 0.040 $\pm$ 0.011	& $1.056^{+0.080}_{-0.029}$\\
$M_*$ [M$_{\odot}$]		& 1.000 $\pm$ 0.065		& 1.021 $\pm$ 0.055 $\pm$ 0.030	& $1.000^{+0.063}_{-0.067}$\\
$\rho_*$ [$\rho_{\odot}$]	& 0.79 $\pm$ 0.10			& 0.803 $\pm$ 0.080 $\pm$ 0.000		& 0.84$^{+0.07}_{-0.15}$\\
$R_\mathrm{p}$ [R$_\mathrm{jup}$]		& 1.167 $\pm$ 0.043 		& 1.171 $\pm$ 0.056 $\pm$ 0.012 & 1.14$^{+0.10}_{-0.04}$\\
$M_\mathrm{p}$ [M$_\mathrm{jup}$]		& 1.568 $\pm$ 0.071		& 1.637 $\pm$ 0.075 $\pm$ 0.033 & 1.555$^{+0.067}_{-0.070}$\\
$\rho_\mathrm{p}$ [$\rho_\mathrm{jup}$]	& 0.92 $\pm$ 0.11			& 1.02 $\pm$ 0.14 $\pm$ 0.01 & 1.05 $\pm$ 0.20\\
\hline
\\
\end{tabular}
\end{center}
\end{table}

%% TTVs induced by other effects
\begin{table}[ht]
\caption{Expected TTVs for WASP-5b induced by effects other than that of an additional perturbing planet or Trojan. T$_{mod}$ denotes the modulation time-scale of the host star. }
\begin{center}
\label{tbl:other_effects} 
\begin{tabular}{lcc}
\\
\hline
effect	& expected TTVs		& reference\\
\hline
the Appregate effect		& $\sim$ 1.5 (T$_{mod}$/11years) [s]	& \citet{2010MNRAS.405.2037W}\\
light travel time effect	& $\lesssim$ 1 [s/3years] \footnotemark[a]  	& e.g., \citet{2005MNRAS.359..567A}\\
exo-moon				& $\lesssim$ 20 [ms] & \citet{2009MNRAS.392..181K, 2010arXiv1007.4500W}\\
tidal dissipation		& $\sim$ 5 [ms year$^{-1}$] \footnotemark[b]  & \citet{2009Natur.460.1098H}\\
orbital precessions due to ...		& 	&\\
\hspace{2em}tidal deformations	& $<$ 130 [ms year$^{-1}$] \footnotemark[c]  & \citet{2007MNRAS.377.1511H,2008ApJ...685..543J}	\\
\hspace{2em}general relativistic effect	& $<$ 2  [ms year$^{-1}$]  \footnotemark[c] & \citet{2007MNRAS.377.1511H,2008ApJ...685..543J}	\\
\hspace{2em}stellar quadrupole moment	& $<$ 0.004  [ms year$^{-1}$] \footnotemark[c] & \citet{2007MNRAS.377.1511H,2008ApJ...685..543J}	\\
\hline
\\
\multicolumn{3}{@{}l@{}}{\hbox to 0pt{\parbox{160mm}{\footnotesize
\par\noindent
\footnotemark[a] The case having a Jovian-mass secondary planet with an orbital period of $\sim$ 3 years, which corresponds to the actual observational period. This produces the RV amplitude of $\sim$21~m~s$^{-1}$ which corresponds to the upper limit placed from the RV data.\\
\footnotemark[b] Adopted a value estimated for OGLE-TR-56b, whose properties are similar to the ones of  the WASP-5b system, as an approximation. \\
\footnotemark[c] Equations of (5), (1), and (3) of \cite{2008ApJ...685..543J} are used for calculating precession rates due to  tidal deformations, the general relativistic effect, and stellar quadrupole moment, respectively, and an equation (23) of \cite{2007MNRAS.377.1511H}  into which the rates are substituted is used for calculating the transit timing delays. Eccentricity of the primary planet of 0.032 is adopted which is 3 $\sigma$ upper limit derived in this work.
}\hss}}
\\
\end{tabular}
\end{center}
\end{table}

%-----------------------------------------------------------
%% figures

%% seven light curves

%% difference in two light curves derived by different corrections

%% binned light curves

%% TTV 

%% Upper limits

%% allowed region

%% allowed region zoom

%% bestfit TTV examples

\end{document}